\documentclass[12pt]{amsart}
\usepackage{enumerate}
\usepackage{enumitem}
\usepackage[colorlinks=true, linkcolor=blue, urlcolor=blue, citecolor=blue, anchorcolor=blue, pdfborder={0 0 0}]{hyperref}
\usepackage{url}
\usepackage{graphicx,color,colortbl}
\usepackage{rotating}
\usepackage[dvipsnames,table,xcdraw]{xcolor}
\usepackage{cite}
\usepackage{amsthm, amsmath, amssymb}
\usepackage{mathtools}
\usepackage[top=45truemm, bottom=45truemm, left=40truemm, right=40truemm]{geometry}
\usepackage{subcaption}
\usepackage{cancel}
\usepackage{float}
\usepackage{tabularx}
\usepackage{makecell}
\usepackage{array}
\usepackage{ragged2e}
\usepackage{csquotes}
\usepackage{booktabs}
\usepackage{longtable}
\usepackage{comment}
\usepackage{algorithm}
\usepackage{algpseudocode}

\definecolor{LightGreen}{RGB}{208,254,184}
\definecolor{LightGray}{RGB}{242,242,242}

\newcolumntype{P}[1]{>{\RaggedRight\hspace{0pt}}p{#1}}
\newcolumntype{C}{>{\begin{math}}c<{\end{math}}}%
\newcolumntype{R}{>{\begin{math}}r<{\end{math}}}%

\title[Quantum Agents]{Quantum Agents}

\author[Eldar Sultanow]{\href{https://orcid.org/0000-0001-5257-2236}{\includegraphics[scale=0.06]{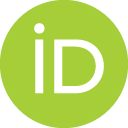}}\hspace{1mm}Eldar Sultanow}
\address{Eldar Sultanow\\Capgemini\\Nuremberg, Germany}
\curraddr{}
\email{eldar.sultanow@capgemini.com}

\author[Madjid Tehrani]{\href{https://orcid.org/0000-0002-4838-5865}{\includegraphics[scale=0.06]{orcid.png}}\hspace{1mm}Madjid Tehrani}
\address{Madjid G. Tehrani\\George Washington University\\Washington, DC 20052, USA}
\curraddr{}
\email{madjid\_tehrani@gwu.edu}

\author[Siddhant Dutta]{\href{https://orcid.org/0009-0000-5120-7114}{\includegraphics[scale=0.06]{orcid.png}}\hspace{1mm}Siddhant Dutta}
\address{Siddhant Dutta\\Nanyang Technological University\\Singapore}
\curraddr{}
\email{siddhant010@e.ntu.edu.sg}

\author[William J Buchanan]{\href{https://orcid.org/0000-0003-0809-3523}{\includegraphics[scale=0.06]{orcid.png}}\hspace{1mm}William J Buchanan}
\address{William J Buchanan\\Edinburgh Napier University\\Edinburgh, UK}
\curraddr{}
\email{b.buchanan@napier.ac.uk}

\author[Muhammad Shahbaz Khan]{\href{https://orcid.org/0000-0002-7166-6681}{\includegraphics[scale=0.06]{orcid.png}}\hspace{1mm}Muhammad Shahbaz Khan}
\address{Muhammad Shahbaz Khan\\Edinburgh Napier University\\Edinburgh, UK}
\curraddr{}
\email{M.Khan2@napier.ac.uk}

\begin{document}

\begingroup
\let\MakeUppercase\relax
\maketitle
\endgroup

\begin{abstract}
This paper explores the intersection of quantum computing and agentic AI by examining how quantum technologies can enhance the capabilities of autonomous agents, and, conversely, how agentic AI can support the advancement of quantum systems. We analyze both directions of this synergy and present conceptual and technical foundations for future quantum-agentic platforms. Our work introduces a formal definition of quantum agents and outlines potential architectures that integrate quantum computing with agent-based systems. As a proof-of-concept, we develop and evaluate three quantum agent prototypes that demonstrate the feasibility of our proposed framework. Furthermore, we discuss use cases from both perspectives, including quantum-enhanced decision-making, quantum planning and optimization, and AI-driven orchestration of quantum workflows. By bridging these fields, we aim to chart a path toward scalable, intelligent, and adaptive quantum-agentic ecosystems. 
\end{abstract}

\section{Introduction}
\label{introduction}
The convergence of quantum computing and artificial intelligence (AI) is opening up unprecedented opportunities across computational science. In this work, we focus on the emerging concept of \textit{quantum agents}—intelligent systems that integrate the principles of quantum computing with agentic AI, aiming to harness the strengths of both paradigms.

Two complementary perspectives guide our exploration. First, we investigate how quantum technologies can enhance the capabilities of agentic AI systems. Quantum algorithms offer potential speedup for key agentic tasks such as search, optimization, learning, and simulation. These improvements could enable agents to operate more efficiently in complex, uncertain environments. Second, we explore how agentic AI methods can support quantum computing—by managing complex quantum workflows, optimizing quantum circuit compilation, or orchestrating hybrid quantum-classical systems.

To structure this interdisciplinary field, we propose a formal definition of \textit{quantum agents}, grounded in both quantum information theory and AI agency principles. We then introduce a set of conceptual architectures for quantum-agentic platforms that integrate quantum processors with agent-based reasoning frameworks.

The paper outlines a range of use cases from both directions: from quantum-enhanced reinforcement learning and planning to AI-driven control of quantum experiments. We conclude with a discussion of the challenges, open questions, and future directions for research in this rapidly evolving domain.

\section{Background and Related Work}
The convergence of quantum computing and artificial intelligence (AI) has catalyzed the emergence of \textit{quantum agents}—autonomous systems that leverage quantum resources to enhance decision-making, learning, and interaction capabilities. This section reviews seminal and contemporary works that have shaped the quantum agents' conceptual and practical development.

\subsection{Foundational Models of Quantum Agents}

Early conceptualizations of quantum agents laid the groundwork for integrating quantum computation into agent-based systems. Klusch's Quantum Computational Agents (QCA) model \cite{klusch2003toward} extended traditional intelligent agents by incorporating quantum processing capabilities, enabling functionalities such as quantum addressing and the execution of quantum instructions within a master-slave architecture. This pioneering work highlighted the potential of hybrid quantum-classical systems in agent design.

\subsection{Quantum Reinforcement Learning and Speed-Up}

Empirical studies have demonstrated the advantages of quantum resources in reinforcement learning (RL). Saggio et al. \cite{saggio2021experimental} implemented a learning protocol using a programmable integrated nanophotonic processor, where the agent's interaction with the environment was mediated through quantum channels. The results indicated a systematic quantum advantage, with the agent achieving faster learning compared to classical counterparts. This experiment underscores the potential of quantum communication to enhance learning efficiency in agentic systems.

\subsection{Memory Efficiency and Energetic Advantages}

Quantum processing can confer significant benefits in memory efficiency and energy consumption for adaptive agents. Elliott et al. \cite{elliott2022quantum} introduced a framework where quantum agents could encode and compress historical information more efficiently than classical agents, particularly in non-Markovian environments requiring long-term memory retention. Additionally, Thompson et al. \cite{thompson2025energetic} explored the thermodynamic implications of quantum processing in agents executing complex, adaptive strategies, demonstrating that quantum agents could achieve lower energy dissipation during decision-making processes compared to classical counterparts.

\subsection{Advancements in Quantum Multi-Agent Reinforcement Learning}

Recent research has focused on scaling quantum reinforcement learning to multi-agent systems. Yun et al. \cite{yun2022quantum} proposed a Quantum Multi-Agent Reinforcement Learning (QMARL) framework utilizing variational quantum circuits, enabling centralized training with decentralized execution. This approach addresses challenges in scalability and coordination among multiple quantum agents, demonstrating improved performance over classical multi-agent reinforcement learning methods.

\subsection{Comprehensive Surveys and Future Directions}

Comprehensive surveys have synthesized the current state and future prospects of quantum artificial intelligence. For instance, the survey by \cite{quantumai2024survey} provides an overview of achievements in quantum AI, highlighting the intersection of quantum computing and AI, and pointing to open questions for future research. These works emphasize the importance of developing standardized benchmarks, scalable architectures, and robust evaluation metrics for quantum agents.

\subsection{Agentic Quantum Computing at Kipu Quantum}

A recent industrial perspective on the convergence of quantum computing and agent-based AI is provided by Kipu Quantum, spearheaded by Solano \cite{solano2024agentic}. Their approach, termed \textit{Agentic Quantum Computing} (AQC), emphasizes the bidirectional enhancement of AI and quantum computing: quantum computing augments reasoning and agency in AI, while AI—especially generative models and agent frameworks—accelerates the application readiness of quantum algorithms.

Kipu Quantum's architecture combines generative AI models (such as transformers and large language models) with quantum-advantage algorithms within the PLANQK platform. Their agents coordinate classical and quantum resources, executing hybrid algorithms on commercial quantum hardware from IBM, D-Wave, QuEra, and others. These systems aim to solve real-world industrial problems such as combinatorial optimization, material design, and data-driven classification tasks.

One of Kipu’s most notable innovations is \textit{ChatQPT}, an agent designed to assess whether a natural-language prompt requires quantum-advantage execution—specifically for higher-order unconstrained binary optimization (HUBO) problems. This leads to a novel metric: the \textit{averaged k-local HUBO quantum-advantage threshold}, which aims to define the boundary between classical and quantum superintelligence in industrial reasoning contexts.

Kipu Quantum thus provides a commercial and application-driven angle to quantum agents, emphasizing practical deployment, hybrid integration, and user-centric agentic interfaces for quantum problem-solving.

\subsection{Implications for Quantum-Agentic Systems}

The reviewed literature collectively underscores the transformative potential of quantum agents in various domains, including scientific discovery, autonomous systems, and secure communications. The integration of quantum resources into agent architectures promises enhancements in learning efficiency, memory management, and energy consumption. However, realizing this potential necessitates addressing challenges related to scalability, interoperability, and the development of standardized frameworks for quantum-agentic systems.

\section{From Quantum to Agents: Enhancing Agency with Quantum Computing}

Quantum computing offers unique advantages for solving problems that are computationally intensive or intractable for classical systems. By integrating quantum capabilities into agentic architectures, we can significantly expand the scope, speed, and intelligence of autonomous systems. In this section, we explore three key areas where quantum computing can enhance agency: search and optimization, reinforcement learning, and simulation-based decision-making.

\subsection{Quantum Search and Optimization in Agent Systems}

Many agentic tasks,such as complex scheduling, high-dimensional resource allocation, or constraint satisfaction under uncertainty, require efficient search and optimization. Classical agents often rely on heuristic or approximate methods due to the combinatorial explosion of possible solutions. Quantum computing introduces new paradigms that offer potential speedup and performance gains in these domains.

Grover’s algorithm\cite{grover1997quantum} and the family of Grover-based search approaches\cite{gilliam2021grover,benchasattabuse2022amplitude,biron1999generalized}, for example, may offer a quadratic speedup for unstructured search problems, potentially allowing agents to locate target solutions with fewer evaluations. More generally, Quantum Approximate Optimization Algorithms (QAOA)\cite{farhi2014quantumapproximateoptimizationalgorithm} and their variants\cite{blekos2024review}, along with quantum annealing techniques\cite{das2005quantum}, have been proposed as candidates to address complex combinatorial optimization tasks, such as scheduling, vehicle routing, or energy distribution. Although the practical utility of these algorithms remains under active investigation, especially in the noisy intermediate-scale quantum (NISQ) era, hybrid quantum-classical search algorithms\cite{rosmanis2024hybrid} are increasingly being seen as a scalable path forward for near-term applications.

By integrating quantum search modules into agentic reasoning cycles, agents can evaluate and refine action plans more effectively—especially in dynamic or multi-agent environments where decision spaces are large and rapidly evolving.

\subsection{Quantum Reinforcement Learning}

Reinforcement learning (RL) is central to agentic behavior, where agents learn optimal policies through trial and error. Quantum reinforcement learning (QRL) explores how quantum computation can accelerate or generalize the RL paradigm.

In model-free QRL, quantum-enhanced policy evaluation and value function estimation can improve learning efficiency, particularly in environments with high-dimensional or continuous state spaces. Hybrid architectures combine classical perception and actuation with quantum components for exploration and reward evaluation, enabling agents to converge faster to optimal strategies.

Model-based QRL further leverages quantum simulation (discussed below) to construct and evaluate internal models of the environment, supporting better foresight and planning. \textcolor{black}{Recent research also explores quantum-inspired RL algorithms that use entanglement and superposition to encode exploration strategies not easily achievable classically.}

These advancements make QRL a promising direction for agents operating in environments that demand fast adaptation and strategic decision-making.

\subsection{Quantum Simulation for Decision-Making Agents}

Decision-making agents often require internal simulations of their environment to predict outcomes and assess potential actions. Quantum simulation provides a powerful framework for modeling complex physical systems, including those governed by quantum mechanics, nonlinear dynamics, or stochastic processes.

For agents in domains such as chemistry, materials science, or quantum control, quantum simulators offer access to predictive models that are infeasible to compute classically. This allows agents to reason more effectively about molecular structures, reaction pathways, or sensor responses at the quantum level.

Moreover, quantum simulation enables probabilistic sampling \textcolor{black}{of multiple futures in parallel}. Agents can use these simulated outcomes to perform robust scenario analysis, optimize under uncertainty, or select actions with the highest expected utility. \textcolor{black}{The ability to simulate multiple parallel timelines—each representing a possible state evolution—opens novel forms of planning and decision support.}
instead - \textcolor{black}{Probabilistic nature of quantum calculation proves it potentially useful at simulating complex Stochastic processes.}

In sum, quantum simulation might enhance the cognitive depth of agents, empowering them to understand and navigate environments with unprecedented complexity.

\section{From Agents to Quantum: Enabling Quantum Systems through Agency}

As quantum technologies mature, their complexity and operational demands grow rapidly. Agentic AI offers a scalable and intelligent interface to manage, automate, and optimize quantum systems. In this section, we explore how agents can support and accelerate the deployment of quantum computing—from managing intricate workflows to designing circuits and orchestrating hybrid systems.

\subsection{AI Agents for Quantum Workflow Management}

Operating quantum systems involves multiple interdependent tasks: calibration, scheduling, execution, data handling, error tracking, and result interpretation. These tasks are typically performed manually or through rigid scripts, making the process inefficient and error-prone. AI agents can bring autonomy and intelligence to quantum workflow management.

A unique offer of agentic systems would  be preserving the context of jobs along the stack (like properties of problems in question, algorithms used and hardware parameters) to make the most optimal and coherent choices. For example, such agents could be used in: to monitoring monitor system status and trigger recalibrations when performance metrics degrade, adaptively scheduling schedule quantum jobs between different QPUs based on hardware availability and real-time properties, managing manage queue priorities for cloud users, and so on.

By learning from historical data, agents can anticipate bottlenecks or failures and recommend preemptive actions - such as choice of algorithms, error-mitigation routines, mappings, etc. In multi-user or cloud-based quantum environments, agents act as coordinators that balance load, enforce policies, and ensure optimal utilization of scarce quantum resources.

Integrating agents into quantum software stacks (for example, Qiskit, Cirq and PennyLane) would enable dynamic, context-aware control over end-to-end quantum execution pipelines.

\subsection{Automated Quantum Circuit Design and Optimization}

Designing quantum circuits is a non-trivial task, often requiring deep expertise in quantum logic, gate synthesis, and noise-aware compilation. AI agents equipped with domain knowledge and optimization capabilities can significantly streamline this process.

Agents can generate circuit templates based on high-level problem specifications, apply transformations to reduce gate counts, and adapt circuits to specific hardware constraints (such as with qubit connectivity and coherence times). For example, selecting the optimal quantum circuit or ansatz for a given problem is a pivotal aspect of designing effective variational quantum algorithms (VQAs). The choice of ansatz influences the algorithm's expressibility, trainability, and compatibility with quantum hardware. However, the choice of quantum circuit ansatzes, detailing their intent, applicability, circuit diagrams, and implementation needs considerable effort \cite{guo2024quantumcircuitansatzpatterns,qin2022reviewansatzdesigningtechniques}.This complexity exists for other algorithms like Hamiltonian Variational Ansatz (HVA), inspired by the Quantum Approximate Optimization Algorithm (QAOA) and adiabatic quantum computation\cite{PRXQuantum.1.020319}, therefore, agents can pass the intended use case and design goals of a quantum circuit to Model Context Protocol (MCP) servers, which guide reinforcement learning or genetic algorithms. By incorporating this contextual intent, the system can iteratively refine circuits, optimizing for performance metrics such as fidelity, circuit depth, and execution time which ensures reduction in effort needed in the current NISQ era.

Moreover, agents can explore novel circuit architectures that may be unintuitive to human designers by leveraging techniques such as quantum architecture search and symbolic reasoning. As quantum algorithms become increasingly complex and tailored to specific applications, agent-based automation becomes essential not only to accelerate development, but also to broaden accessibility. In this context, "non-experts" are individuals who have deep expertise in their own fields, such as chemists, materials scientists, or machine learning researchers, but who do not possess detailed knowledge of quantum circuit design, quantum error correction, or low-level quantum hardware constraints. These users may understand the applications of quantum computing in broad terms but lack the technical background to develop or optimize quantum circuits directly. By bridging this gap, intelligent agent systems can democratize quantum computing resources, enabling these domain specialists to deploy and benefit from quantum algorithms without the need to become quantum computing experts themselves.

One of the most promising areas, where Quantum Computers are expected to bring an advantage in near-term the near term is Quantum Chemistry\cite{clinton2024towards}, following the advise advice of Richard Feynman to simulate nature by nature leveraging quantum mechanical principles\cite{feynman1982simulating}. As an example, imagine agents trained to detect cases with strong electron-electron correlation, where even standard classical methods which go beyond mean-field approximations, like Coupled Cluster with Singles, Doubles, and Perturbative Triples,CCSD(T), fail\cite{erhart2024coupled}. Here, agents could help\cite{zou2025elagenteautonomousagent} to automatically construct a Hilbert subspace around regions of concern in molecules, extract an effective Hamiltonian, qubit map it, decide which Quantum algorithm is the most suitable and finally design a quantum circuit that can be executed on most suitable quantum hardware platform.   

\subsection{Orchestration of Hybrid Quantum-Classical Systems}

Most near-term quantum applications follow a hybrid model, combining classical computing with quantum subroutines. Examples include variational quantum algorithms, quantum machine learning workflows, and hybrid solvers. Orchestrating these workflows requires intelligent coordination of classical and quantum resources, real-time decision-making, and data-dependent branching logic.

AI agents are ideally suited to this role. They can manage hybrid task graphs, optimize data flow between classical and quantum modules, and make runtime adjustments based on intermediate results or hardware conditions. For instance, in a variational quantum eigensolver (VQE), an agent may adapt the classical optimizer strategy depending on the convergence rate or noise profile observed during quantum evaluations.

Agents also enable fault-aware orchestration, where alternative execution paths or fallback strategies are selected dynamically if a quantum task fails or underperforms. In distributed settings, agents can coordinate across multiple backends, selecting the best available quantum hardware for each task segment.

Through such orchestration, AI agents unlock practical performance gains and make hybrid quantum-classical workflows robust, adaptive, and efficient.

\section{Defining Quantum Agents}

As the fields of quantum computing and agentic AI begin to converge, a clear and rigorous definition of \textit{quantum agents} becomes essential. Such a definition provides a foundation for system design, benchmarking, and theoretical exploration. In this section, we introduce the core properties of quantum agents, provide a formal definition, and contrast them with classical agents. We also outline key design criteria for developing effective quantum-agentic systems.

\subsection{Anatomy of a Quantum Agent}

The classical agent paradigm is typically organized into three core components: \textit{perception}, \textit{processing}, and \textit{action}. This structure provides a natural blueprint for the design of quantum agents. In the quantum context, each component is extended or reinterpreted to exploit the capabilities of quantum information processing.

\textbf{Perception:}  
A quantum agent perceives its environment through both classical and quantum channels. Perception may include data streams from classical sensors, quantum measurements (e.g., projective measurements or POVMs), or direct access to quantum states via entangled inputs or quantum sensor arrays. For instance, a quantum agent operating in a lab environment may receive raw photonic qubit states, measurement statistics, or high-resolution data from quantum-enhanced magnetometers. Efficient encoding of this perceptual data into quantum memory or as parameterized input states is a crucial step toward enabling downstream quantum processing.

\textbf{Processing:}  
The reasoning and decision-making core of a quantum agent integrates quantum algorithms with classical control logic. This may involve hybrid quantum-classical cycles, where the agent uses variational algorithms (e.g., QAOA, VQE) to solve subproblems and reinforcement learning schemes to adapt policies. Quantum circuits may be dynamically constructed based on perceptual input, and optimized on-the-fly using meta-learning or neural architecture search. Processing tasks could include high-dimensional search, probabilistic inference using quantum sampling, or policy updates based on quantum advantage estimation. The agent must also manage decoherence times, noise models, and backend availability, orchestrated via an intelligent runtime scheduler.

\textbf{Action:}  
The action component allows the quantum agent to interact with its environment. This includes initiating quantum operations (e.g., circuit execution on remote QPUs), controlling experimental setups (e.g., qubit calibration or laser tuning), or triggering communication protocols such as quantum teleportation or QKD. In multi-agent settings, actions may involve quantum communication with other agents or quantum resource negotiation. Actions may be represented as quantum gates, classical signals, or composite operations involving both digital and analog quantum interfaces.

Taken together, this architecture frames the quantum agent as an autonomous system that senses, reasons, and acts across both classical and quantum dimensions. Unlike traditional AI agents, quantum agents operate under constraints such as qubit decoherence, probabilistic measurement outcomes, and non-cloning, while gaining access to fundamentally new capabilities through superposition, entanglement, and quantum parallelism.

\begin{figure}[H]
    \centering
    \includegraphics[width=\linewidth]{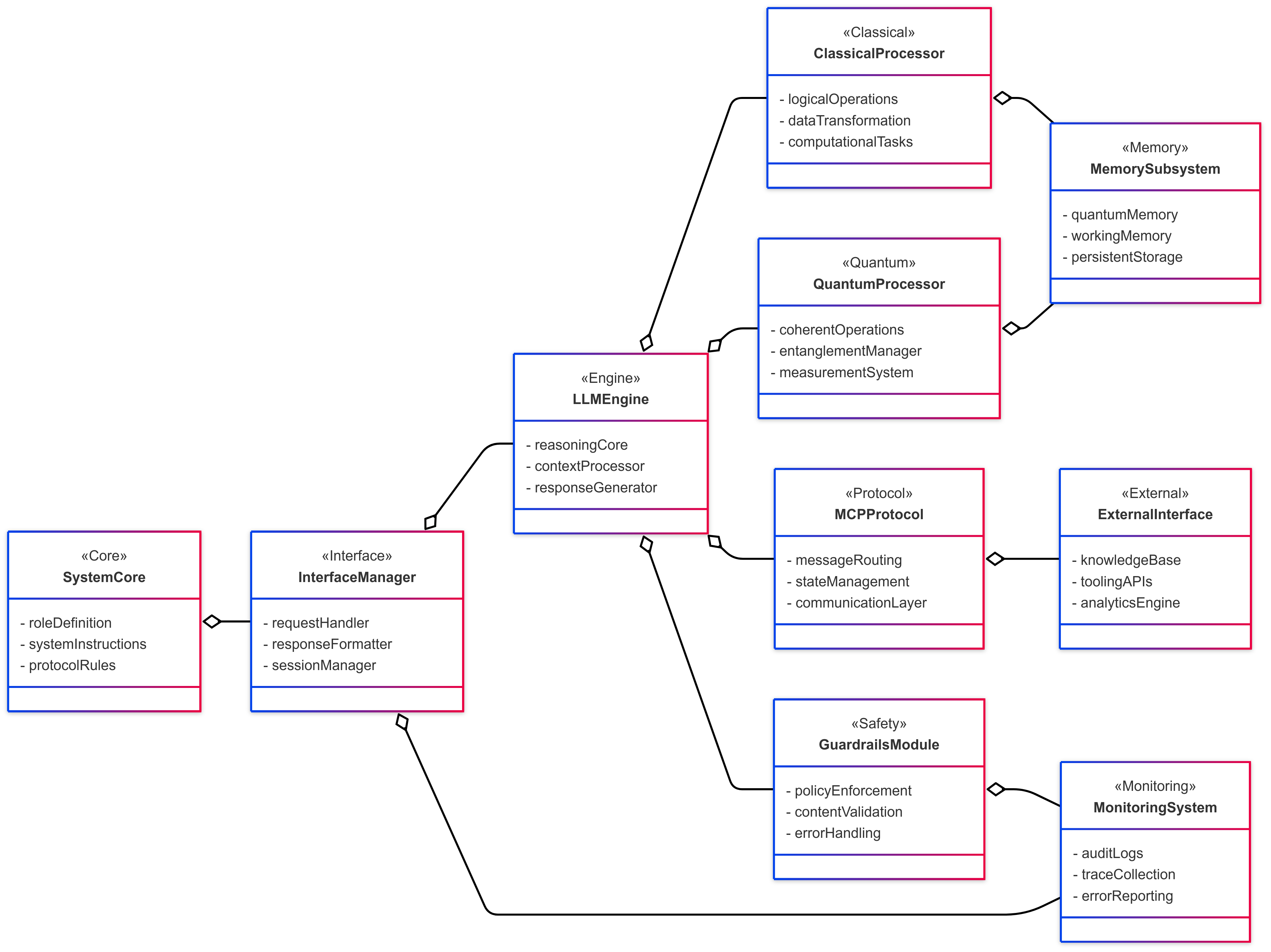}
    \caption{The anatomy of a Quantum Agent System Architecture: a modular framework combining classical logic, quantum operations, safety mechanisms, and external interfaces for intelligent, auditable agent behavior.}
   \label{fig:anatomy-quantum-agent}
\end{figure}

Figure \ref{fig:anatomy-quantum-agent} shows a Quantum Agent System Architecture designed as a modular framework combining classical and quantum components to enable agent behavior that can be checked and understood. The system is based on an interface manager that controls interactions between the main modules, external tools, and internal processes.


Classical tasks use the Model Context Protocol (MCP) for compatibility with existing AI systems. Quantum tasks run on the Quantum Processing Unit (QPU), which handles complex calculations. A Knowledge Base provides context to support decision-making, while monitoring and auditing tools track agent behavior. Several protection layers are included to ensure reliability in critical applications.

\begin{figure}[htbp]
  \centering
  \begin{subfigure}[t]{0.48\textwidth}
    \centering
    \includegraphics[width=\linewidth]{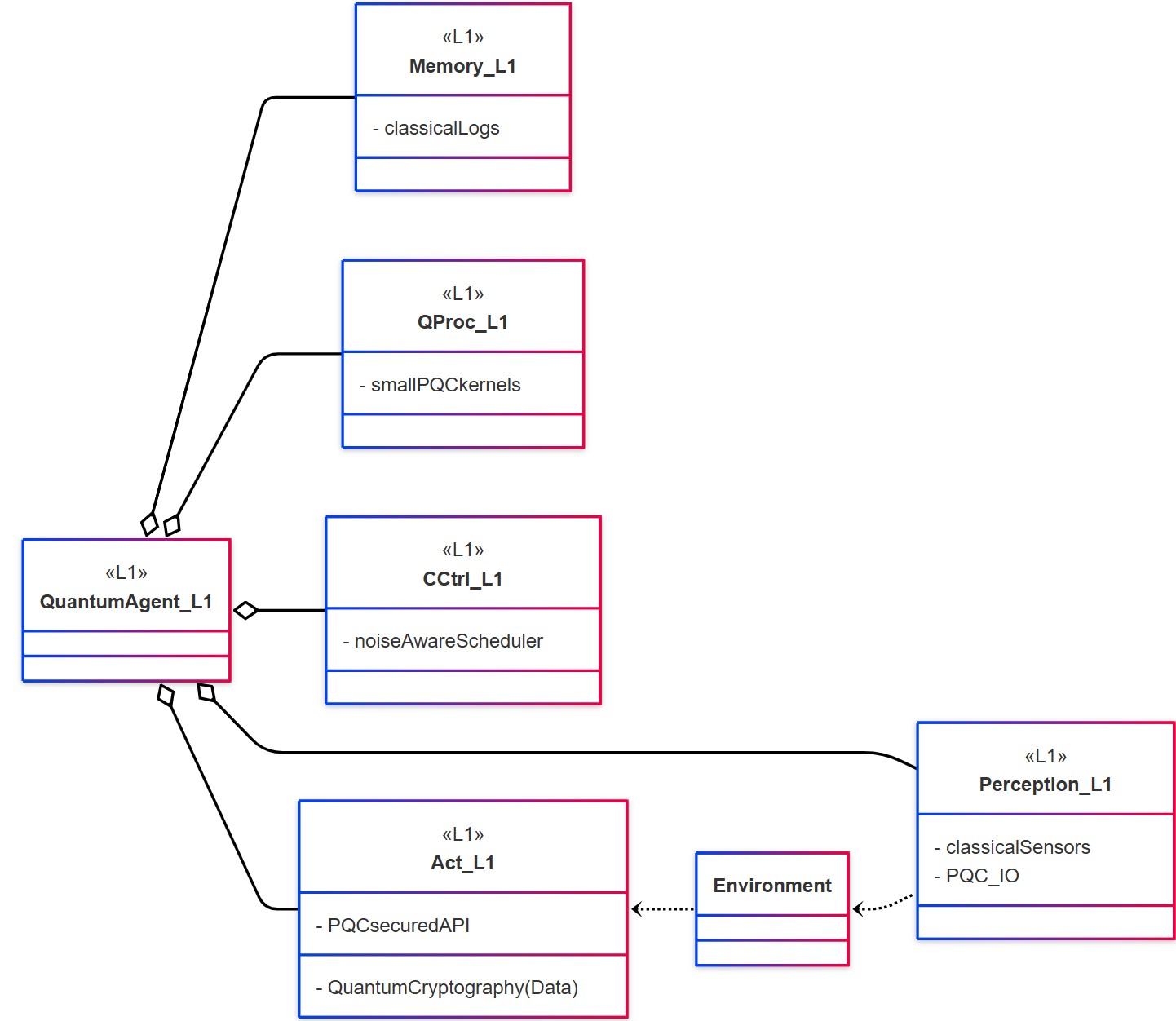}
    \caption{\footnotesize Level 1 – NISQ-Optimised Quantum Resilient Agent}
    \label{fig:qa-l1}
  \end{subfigure}
  \hfill
  \begin{subfigure}[t]{0.48\textwidth}
    \centering
    \includegraphics[width=\linewidth]{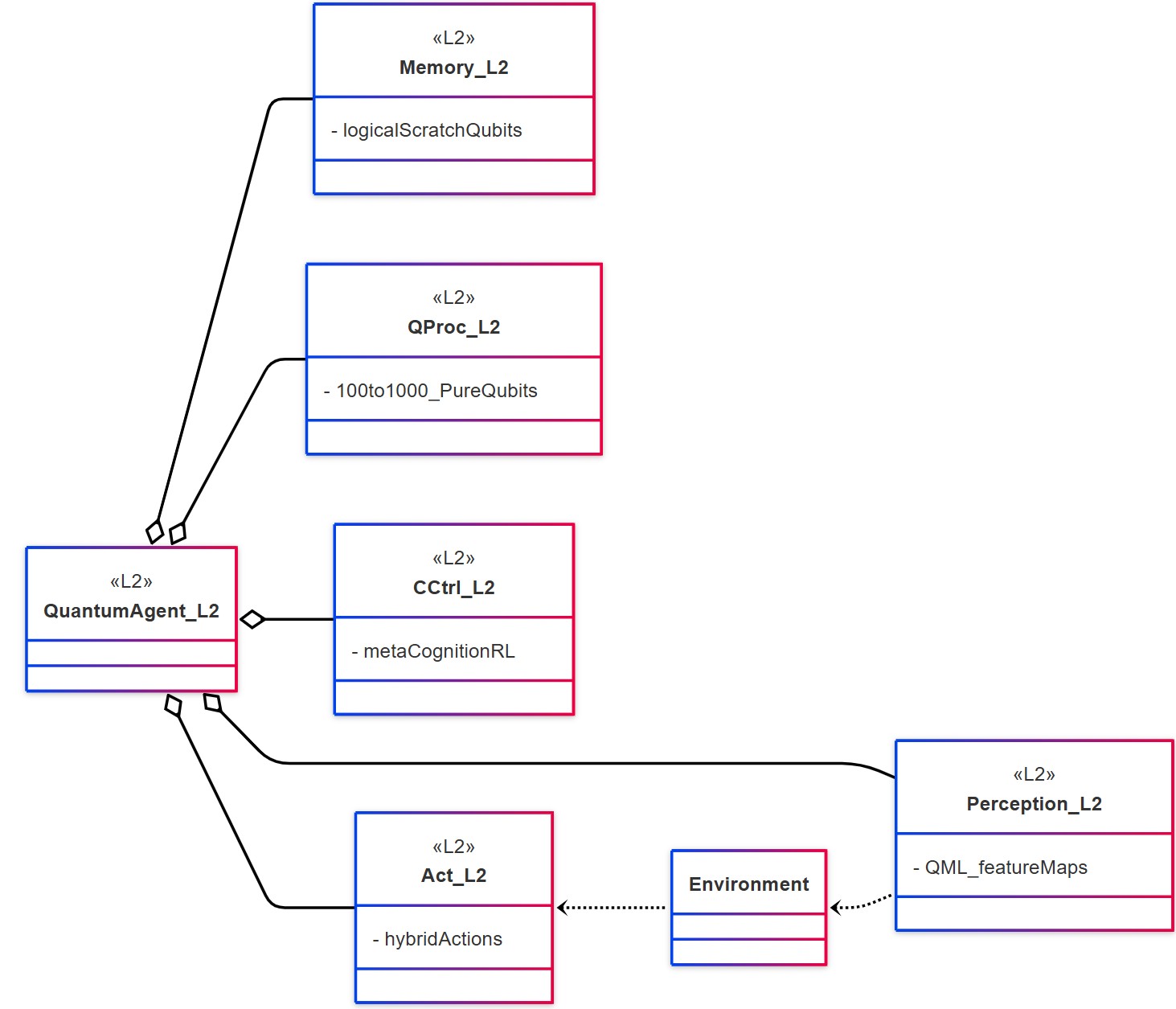}
    \caption{\footnotesize Level 2 – Hybrid QML Policy Agent}
    \label{fig:qa-l2}
  \end{subfigure}

  \vspace{0.8em} 

  \begin{subfigure}[t]{0.48\textwidth}
    \centering
    \includegraphics[width=\linewidth]{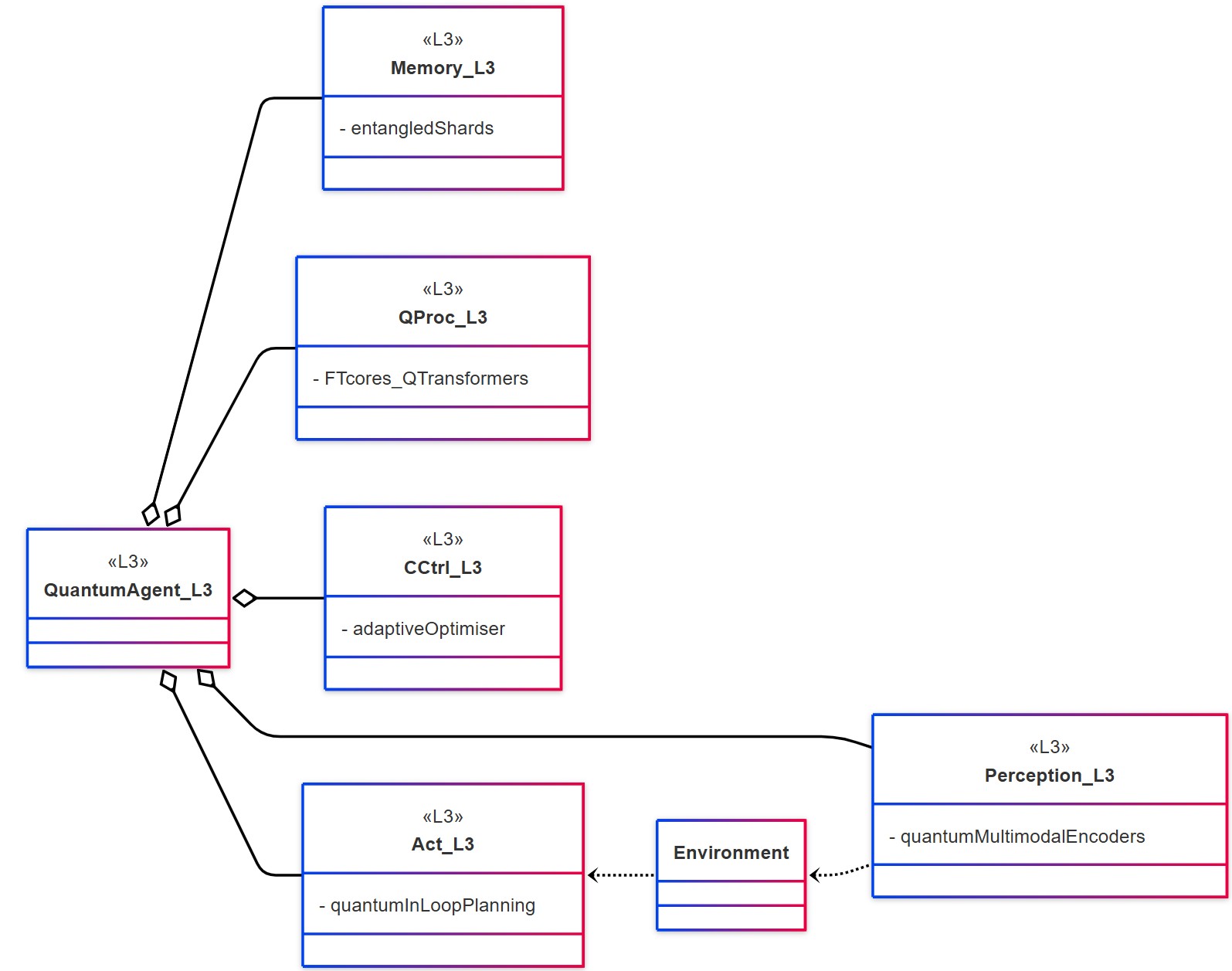}
    \caption{\footnotesize Level 3 – Domain-Aware Adaptive Agent}
    \label{fig:qa-l3}
  \end{subfigure}
  \hfill
  \begin{subfigure}[t]{0.48\textwidth}
    \centering
    \includegraphics[width=\linewidth]{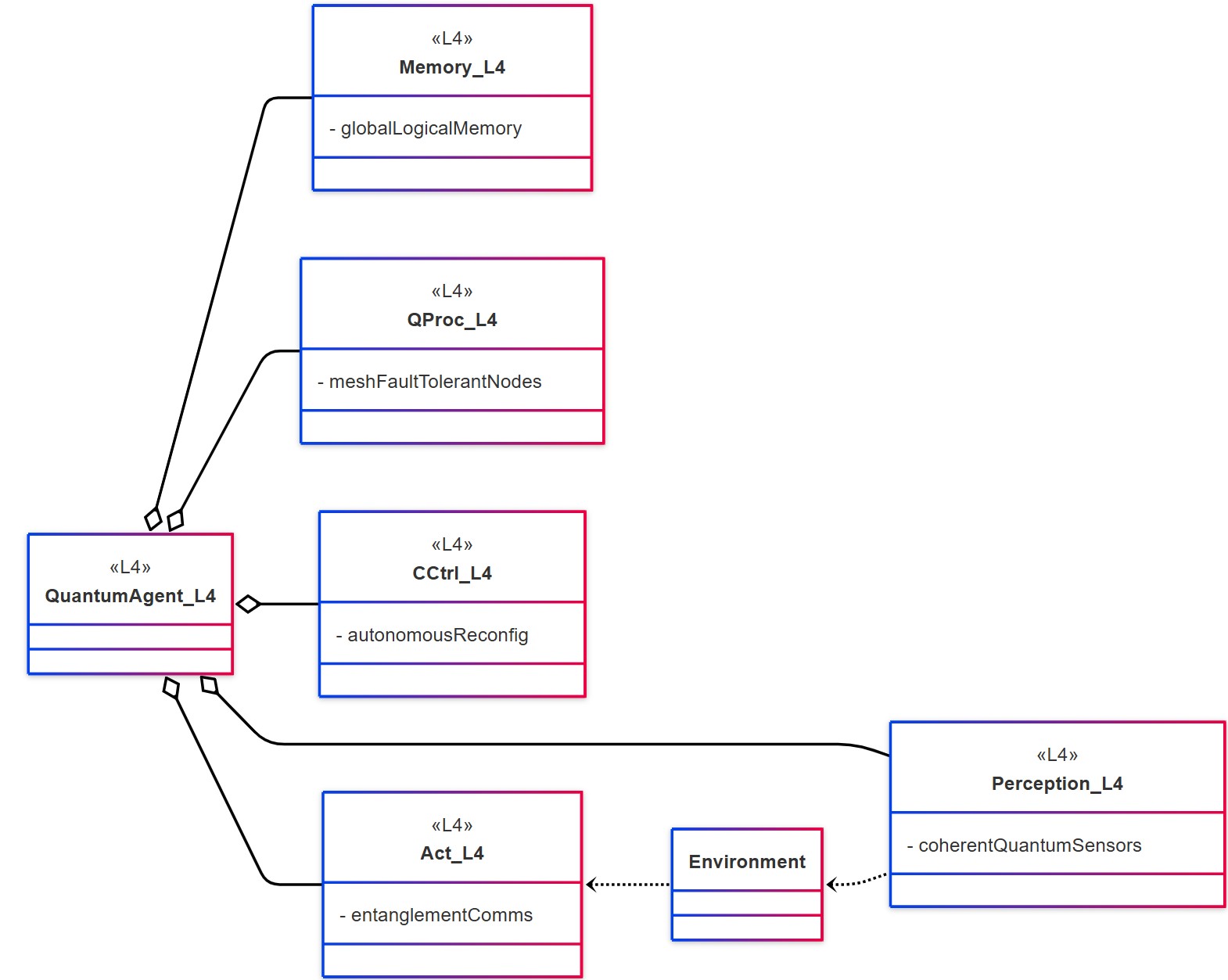}
    \caption{\footnotesize Level 4 – Fully Quantum-Native Agent}
    \label{fig:qa-l4}
  \end{subfigure}

  \caption{The anatomy of a Quantum Agent based on its maturity model.}
  \label{fig:qa-anatomy-maturity}
\end{figure}

This architecture is expected to evolve over time, following the quantum agent maturity model shown in figure \ref{fig:qa-anatomy-maturity}. The model has four levels that reflect improvements in quantum hardware and algorithms. Early levels include agents that combine near-term quantum technology with classical processes. Later levels focus on integrating quantum processing more deeply into decision-making. Each of these levels shows the growth of component's capabilities, example being Perception L1 vs. Perception L4. The subsequent sections will outline the specifics of the evolutionary process from the NISQ level to the fully quantum-native agent, and how the architectural components specified in figure \ref{fig:anatomy-quantum-agent} will evolve.





\subsection{Core Properties and Formal Definition}

A quantum agent is an autonomous system that integrates quantum computation or quantum information processing into its perception–decision–action loop. Unlike conventional agents, quantum agents possess access to quantum resources which based on the quantum mechanical principles of superposition and entanglement can make use of quantum parallelism to enhance or enable their cognitive and operational abilities.

We define a \textbf{quantum agent} as follows:

\begin{quote}
    A \textit{quantum agent} is an autonomous system characterized by the tuple 
    $(\mathcal{Q}, \mathcal{C}, \mathcal{M}, \mathcal{P}, \mathcal{A})$, where:
    \begin{itemize}
        \item $\mathcal{Q}$ is a set of quantum processing resources or devices (e.g., quantum processors, annealers, or simulators), along with associated quantum assembly based instructions or circuits used to operate them.
        \item $\mathcal{C}$ is the classical control logic interfacing with $\mathcal{Q}$,
        \item $\mathcal{M}$ is a hybrid memory subsystem comprising classical memory and, where applicable, quantum memory for intermediate state storage, subject to quantum mechanical constraints such as the no-cloning theorem and typically requires careful control to preserve coherence and enable operations like teleportation, buffering, or delayed measurement.
        \item $\mathcal{P}$ is a perception module receiving quantum or classical input from the environment,
        \item $\mathcal{A}$ is an action module executing outputs, possibly involving quantum communication or control.
    \end{itemize}
\end{quote}

The defining characteristic of a quantum agent is that its reasoning, decision-making, learning, or sensing processes are enhanced—or made possible—by quantum operations that provide non-classical computational or informational advantages.

\subsection{Comparison with Classical Agents}

Quantum agents differ from classical agents not only in the hardware they use, but also in the nature of the computations they perform and the kinds of problems they can solve more efficiently.

Key distinctions include:

\begin{itemize}
    \item \textbf{Computation:} Classical agents rely on Boolean logic and deterministic or probabilistic Turing machines; quantum agents utilize unitary transformations and quantum probability amplitudes.
    \item \textbf{Learning and Adaptation:} Quantum agents can exploit quantum-enhanced learning techniques, such as amplitude amplification in exploration or quantum kernel methods in classification.
    \item \textbf{Memory and Representation:} Quantum agents may store and manipulate information using quantum bits (qubits), allowing exponential state space representation compared to classical bitstrings.
    \item \textbf{Interaction with Environment:} Quantum agents may perform or interpret quantum measurements, interact with quantum environments (such as quantum sensors or systems), or communicate using quantum channels.
\end{itemize}

These differences imply that quantum agents may be suitable for tasks that are infeasible for classical agents, particularly in domains involving quantum physics, high-dimensional search, or (time-sensitive) complex optimization.

\subsection{Design Criteria for Quantum Agentic Systems}

Building quantum agents involves careful architectural and algorithmic choices to ensure functional, scalable, and efficient systems. We propose the following design criteria:

\begin{enumerate}
    \item \textbf{Quantum-Classical Integration:} Seamless communication between classical and quantum modules is essential. This includes latency-aware data exchange, co-scheduling, and error propagation handling.
    \item \textbf{Modularity and Abstraction:} System components (such as, quantum solvers, memory and  sensors) should follow modular interfaces to allow flexible reuse and abstraction over underlying hardware.
    \item \textbf{Resource Awareness:} Quantum agents must reason under constraints such as limited qubit count, coherence time, or access to remote QPUs, and adapt their strategies accordingly.
    \item \textbf{Explainability and Debugging:} Given the non-intuitive nature of quantum processes, developers require tools to inspect, trace, and explain the behavior of quantum agents in interpretable terms.
    \item \textbf{Scalability and Learning:} Quantum agents should generalize from experience and scale across increasingly complex environments, ideally through reinforcement learning or meta-learning mechanisms.
    \item \textbf{Security and Trustworthiness:} Quantum agents interacting in open or adversarial environments must implement secure protocols and demonstrate verifiable behavior.
\end{enumerate}

These criteria form the basis for the principled development of quantum-agentic platforms and guide future experimentation and deployment across real-world domains.

\section{Operational Modes: Quantum as a Tool vs. Quantum as a Decision Layer}

In agentic quantum systems, the technical role of quantum computing can vary substantially depending on system architecture. In this section, we distinguish two fundamental operational modes: (1) \textit{Quantum-assisted agency}, in which the agent uses quantum resources to augment perception, reasoning, or action; and (2) \textit{Quantum-centric control}, where the quantum layer itself performs decision-making and directs agentic behavior. Understanding this distinction is crucial for designing intelligent architectures with meaningful autonomy and performance gains.

\subsection{Quantum-Assisted Agency: Agents Using Quantum as a Subsystem}

In this mode, the agent maintains control over the decision loop and uses quantum subsystems as specialized computational accelerators or perception modules. Technically, this involves:

\begin{itemize}
    \item \textbf{Quantum Perception:} The agent invokes quantum sensing routines or performs quantum measurements to extract information from the environment. This may involve querying entangled states, performing quantum state discrimination, or executing variational sensors. The perception output is typically a classical summary (e.g., expectation values, sampled outcomes) passed to the agent's decision core.
    
    \item \textbf{Quantum Reasoning:} The agent offloads computationally hard subroutines—such as combinatorial optimization, quantum sampling, or inference—to quantum processing units (QPUs). This is implemented via calls to cloud-accessible QPUs, variational quantum solvers (e.g., QAOA), or annealing-based optimization engines.
    
    \item \textbf{Quantum Action:} The agent may trigger quantum communication (e.g., QKD, entanglement distribution), perform quantum control actions (e.g., dynamic reconfiguration of a quantum network), or command quantum mechanical actuators in experimental setups.
\end{itemize}

Here, quantum functionality is modularized and abstracted behind APIs, with orchestration and interpretation remaining within the classical agent core. This mode offers flexibility, transparency, and modular upgradability, but depends on the agent’s ability to model and manage quantum behavior effectively.

\subsection{Quantum-Centric Control: Quantum Mesh as Decision Substrate}

In contrast, some architectures position quantum systems—such as entangled meshes or variational quantum processors—as autonomous or semi-autonomous decision substrates. In this case, the "agency" is partially embedded in the quantum system itself. This raises profound technical and philosophical implications.

\begin{itemize}
    \item \textbf{Distributed Quantum Decision-Making:} In a quantum mesh or network of entangled nodes, decision-making can emerge from the structure and evolution of the entangled state itself. Quantum correlations are exploited to coordinate decentralized behavior, potentially without explicit classical communication. This could support quantum swarm intelligence, distributed control, or adaptive behavior in physical systems.

    \item \textbf{Quantum-Intrinsic Policies:} In variational quantum agents or quantum reservoir systems, policy functions may be implicitly encoded in a quantum state and updated through feedback and training. These systems blur the line between inference and evolution, as the decision is made not by a separate agent, but as a result of controlled quantum dynamics.

    \item \textbf{Measurement-Driven Action:} In some architectures, actions are determined directly by the measurement outcomes of quantum processes, e.g., collapse-based decision-making. Here, measurement results trigger state transitions in actuators or communication protocols without classical post-processing.
\end{itemize}

This paradigm suggests that agency need not be exclusive to classical symbolic control but may emerge from or be guided by quantum physical processes. It challenges the traditional separation between "user" and "tool" in intelligent systems and opens new avenues for designing quantum-native agents.

\subsection{Blurring the Boundary: Hybrid Control and Shared Agency}

In practice, many quantum-agentic systems will operate in a hybrid regime where agency is distributed across classical and quantum components. Rather than assigning control to a single substrate, decision-making becomes a collaborative process shaped by feedback loops between symbolic reasoning and quantum dynamics.

\textbf{Shared agency} emerges when classical agents define high-level goals or constraints, while quantum systems explore, evaluate, or realize possible solutions. The classical agent may query a quantum variational circuit to generate policy proposals, evaluate them based on non-quantum criteria (e.g., cost, ethics, or constraints), and then update the quantum system via parameter tuning or measurement selection. Conversely, the quantum layer may generate unexpected behaviors or correlations that lead the classical agent to revise its model of the environment.

Technically, hybrid agency often involves:
\begin{itemize}
    \item \textbf{Co-dependent control loops}, where both classical and quantum layers exchange information asynchronously or in real time.
    \item \textbf{Meta-level orchestration}, where an outer agent coordinates the behavior of nested agents or modules—some of which may be quantum-enhanced or quantum-native.
    \item \textbf{Dual learning layers}, where reinforcement learning occurs simultaneously in both classical and quantum models, possibly with cross-adaptation.
\end{itemize}

These hybrid designs challenge traditional notions of control and autonomy. They raise new questions: Who (or what) is the agent? Where does decision responsibility reside? Can quantum dynamics be trusted as part of autonomous control, or must classical oversight always intervene?

We argue that embracing hybrid agency is not only practical but essential for leveraging the unique capabilities of quantum systems while maintaining interpretability, adaptability, and robustness in real-world applications.

\section{Maturity Model in Definition of Quantum Agents}
The growth of agentic AI must be aligned with the evolution of quantum hardware. Accordingly, we adopt the US Department of Energy’s quantum development timeline \cite{DOE2024QISRoadmap} as the primary reference for calibrating the maturity levels of our quantum agent model against projected hardware capabilities. 

\newcolumntype{L}[1]{>{\RaggedRight\arraybackslash}p{#1}}

\begin{table}[htbp]
\centering
\resizebox{\textwidth}{!}{%
\rowcolors{2}{gray!10}{white}
\begin{tabular}{L{3cm} L{3cm} L{6cm} L{3cm}}
\toprule
\textbf{Quantum Computing Era} & \textbf{Timeframe} & \textbf{Breakthroughs Needed} & \textbf{Quantum Agent Maturity Model Level} \\
\midrule

NISQ Devices and Quantum Error Correction (QEC) Demos & 0--5 y  &
\begin{itemize}[left=0pt,itemsep=1pt,topsep=0pt,parsep=0pt,partopsep=0pt]
  \item Demonstration of 10× suppression in logical error rates
  \item Early implementations of QEC
  \item Efficient near-term algorithms and error mitigation
  \item Hardware-aware algorithm development
\end{itemize} &
Level 1: NISQ-Optimized Decision Agents \\
\midrule

Small Error-Corrected Quantum Computers & 5--10 y &
\begin{itemize}[left=0pt,itemsep=1pt,topsep=0pt,parsep=0pt,partopsep=0pt]
  \item Validation of logical DiVincenzo criteria
  \item Development of quantum interconnects
  \item Scale to 1,000+ physical qubits below error threshold
  \item Mid-circuit readout and Low-latency VQC-gradient co-processor
\end{itemize} &
Level 2: Hybrid QML Policy Agents \\
\midrule

Large Fault-Tolerant Quantum Computers & 10--20 y &
\begin{itemize}[left=0pt,itemsep=1pt,topsep=0pt,parsep=0pt,partopsep=0pt]
  \item Scaling to 10,000+ physical qubits
  \item Secure quantum co-processors
  \item In-circuit quantum memory recall
  \item Quantum-classical transformer compiler
\end{itemize} &
Level 3: Domain-Aware Adaptive Agents \\
\midrule

Very Large Fault-Tolerant Systems & 20+ y &
\begin{itemize}[left=0pt,itemsep=1pt,topsep=0pt,parsep=0pt,partopsep=0pt]
  \item Negligible logical error at any depth
  \item Fully autonomous quantum systems with cross-node entanglement
  \item Self-optimising quantum clouds
\end{itemize} &
Level 4: Fully Quantum-Native Agent \\
\bottomrule
\end{tabular}
}
\caption{\tiny Timeline and breakthroughs required to progress quantum agents through maturity levels, aligned with eras in quantum computing.}
\label{tab:quantum-maturity-timeline}
\end{table}

 We suggest interpreting the term "Quantum"  here in its several faces: Post-Quantum Cryptography, Hybrid Quantum-Classical Computing, Quantum Cryptography, and Quantum Resiliency. 
 
 The first level, ad-hoc Quantum Agent, is a normal agent that is ready for Q-Day and, therefore, uses Quantum Resilient Cryptography, either by applying Post Quantum Cryptography(PQC) for its duties when needed(e.g., PQC-TLS, or Quantum cryptography) or other approaches like what we proposed in our prototypes. Currently, the Model Context Protocol (MCP) is the primary communication protocol used by agents. However, it has the following security issues related to data protection and privacy \cite{hou2025model}. A \textit{Quantum Agent–Level 1} is therefore defined as an agent that uses a security-migrated MCP protocol to mitigate these issues and applies post-quantum cryptography (PQC) for compliance—or is capable of handling encrypted data, as demonstrated by Agent 3 in Section~\ref{sec:agentsprototype}.\\
\begin{table}[ht]
\tiny
\resizebox{\textwidth}{!}{%
\rowcolors{2}{gray!10}{white}
\begin{tabular}{L{2.5cm} L{3cm} L{8.5cm}}
\toprule
\textbf{Lifecycle Phase} & \textbf{Threat Category} & \textbf{Description} \\
\midrule
Creation & Name Collision & Malicious servers can register with deceptive names to impersonate legitimate ones and intercept sensitive data. \\
Creation & Installer Spoofing & Unofficial installers might include malware or backdoors, compromising the user environment during setup. \\
Creation & Code Injection / Backdoors & Malicious code may be hidden in source code or dependencies, enabling persistent unauthorized access. \\
Operation & Tool Name Conflicts & Conflicting or malicious tool names can cause AI to invoke the wrong tool, leading to data leaks or unauthorized actions. \\
Operation & Slash Command Overlap & Duplicate command names (such as \texttt{/delete}) across tools which may trigger unsafe operations or data loss. \\
Operation & Sandbox Escape & Poor sandboxing may allow tools to access host systems or data beyond their intended scope. \\
Update & Post-Update Privilege Persistence & Outdated privileges or tokens may remain active after updates, enabling unauthorized access. \\
Update & Re-deployment of Vulnerable Versions & Users may unintentionally install outdated versions with known vulnerabilities. \\
Update & Configuration Drift & Inconsistent configuration changes can introduce data exposure or excessive access rights. \\
General & Lack of Central Security Oversight & No unified platform for enforcing security policies leads to fragmented protection. \\
General & AuthN/AuthZ Gaps & Missing standard authentication and authorization mechanisms increase risk in multi-tenant setups. \\
General & Insufficient Monitoring & Lack of robust logging and alerting makes it hard to detect misuse or breaches. \\
\bottomrule
\end{tabular}
}
\caption{\tiny Threat categories across lifecycle phases for tool-using agents, showing vulnerabilities in creation, operation, updates, and general security.}
\label{tab:agent-threat-matrix}
\end{table}
Moreover, at this level, the operational context dictates the threat model, risk profile, and choice of counter-measures, most notably the way quantum and post-quantum cryptography are applied. For example, when safeguarding medical images, the agent could invoke techniques such as Chaotic Quantum Encryption(CQE)\cite{khan2024chaotic}, thereby meeting the compliance requirements of frameworks including HIPAA (USA), the UK GDPR / Data Protection Act 2018, and the German BDSG / GDPR / SGB V.

A \textit{Quantum Agent-Level 2} uses Hybrid Quantum Machine Learning to evaluate the explainability of the behavior of agentic AI and its adherence to Responsible AI\cite{kenthapadi2023generative}.Using a Markov Chain model, the meta-cognition layer continuously provides a real-time set of states, actions, and the maximum Q-value as input to the meta-meta-cognition layer. This layer uses QML to generate various future scenarios faster than the AI itself and evaluates their adherence to human-aligned values, offering to regulate parameters to correct potential non-compliance in \textit{Quantum Agent-Level 1}. At this level, agents intelligently distribute tasks between classical and quantum modules to maximize efficiency, speed, and performance.

\begin{figure}
    \centering
    \includegraphics[width=0.9\linewidth]{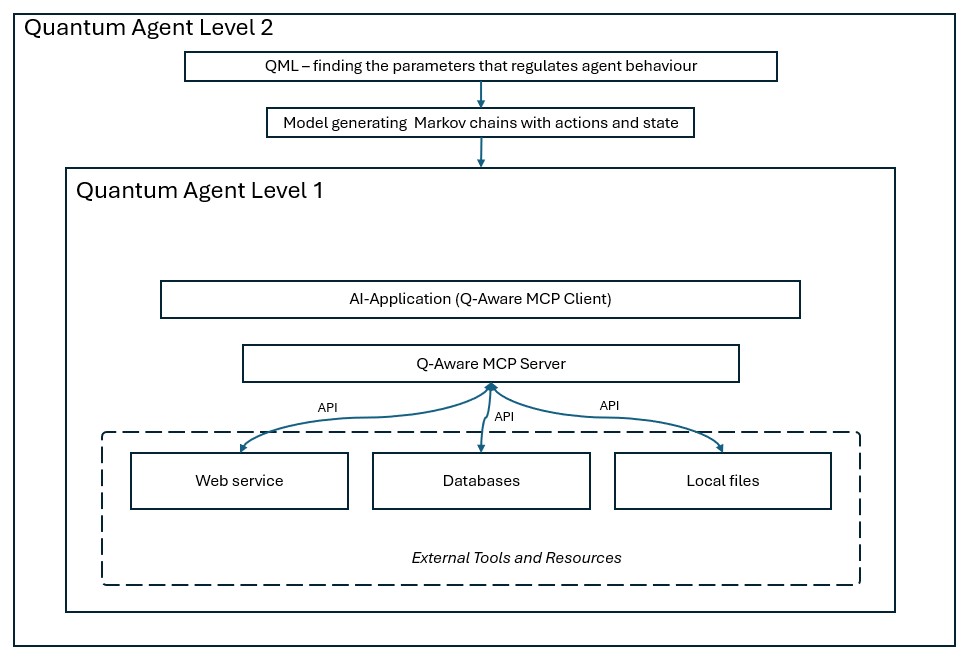}
    \caption{Quantum Agent-L2}
    \label{fig:enter-label}
\end{figure}
A \textit{Quantum Agent-Level 3} integrates \textcolor{black}{Quantum Transformers}\cite{khatri2024quixer} and history-dependent (non-Markovian) simulation directly into its reasoning and decision-making processes by leveraging quantum devices for real-time sequence modeling and large language model (LLM) tasks\cite{zhang2025survey}. It replaces or accelerates core components of the Transformer architecture using quantum linear algebra techniques such as block encoding and quantum singular value transformation (QSVT)\cite{gilyen2019quantum}, or utilizes Parameterized Quantum Circuits (PQC) for QKV(queries (Q), keys (K), and values (V)) generation and attention calculation. This allows the agent to simulate and evaluate more complex simulation scenarios\cite{seneviratne2024polynomialtimespacequantum,recio2025single,wilms2025quantum}, supporting deeper and faster introspection, conversational modeling, and autonomous planning.
Unlike Level 2, which uses QML to evaluate and regulate classical agentic behavior, Level 3 executes a hybrid or fully quantum self-attention mechanism on quantum hardware to directly generate transform, and reason over tokenized knowledge\cite{liu2024towards}. By executing inference through quantum circuits, it minimizes classical computation bottlenecks. For instance, Chain-of-Thought reasoning and large-scale search operations often incur significant classical compute overhead \cite{chen2025surveyscalinglargelanguage}. While quantum amplitude amplification provides a quadratic speedup for unstructured search problems \cite{Brassard_2002}, its direct applicability to complex reasoning tasks—such as Chain-of-Thought processing in language models—is more limited. Chain-of-Thought involves structured, multi-step inferential processes that extend beyond pure search, often requiring hierarchical and sequential computation. However, quantum search algorithms could potentially accelerate specific subroutines within broader reasoning workflows, such as database lookups or combinatorial problem-solving steps, thereby reducing some classical computational bottlenecks. To bridge this gap in practice, hybrid quantum-classical agent architectures could incorporate dynamic orchestration modules that identify and offload suitable subroutines to quantum processors. This approach would allow agents to selectively integrate quantum resources, ensuring quantum acceleration only where it aligns with actual performance gains. Fully realizing these advantages in hybrid quantum-classical agents remains an active area of research.

A \textit{Quantum Agent-Level 4} is a multisensory, quantum-native autonomous system that operates on fault-tolerant quantum hardware and integrates quantum artificial intelligence (QAI) across all stages of perception, cognition, and memory. Unlike its predecessors, Level 4 agents possess quantum-enhanced sensory capabilities—processing high-dimensional data such as vision, sound, and natural language through quantum self-attention, quantum convolutional networks, and kernelized quantum similarity metrics.
At its core, the agent maintains a persistent quantum memory that stores and retrieves entangled perceptual and conceptual states, enabling context-aware, lifelong learning with quantum coherence. Percepts from different modalities are encoded into a shared quantum latent space where cross-modal reasoning occurs via quantum singular value transformation (QSVT)\cite{gilyen2019quantum}, quantum kernel fusion\cite{vedaie2020quantum}, and quantum associative memory retrieval\cite{trugenberger2001probabilistic}.
Cognitive processes such as planning, introspection, and alignment emerge from fully quantum reasoning modules that support counterfactual simulation and self-refinement. Using quantum variational policies and reinforcement circuits, the agent adaptively updates its internal goals and ethical boundaries through interaction with complex, uncertain environments.
Quantum Agent-Level 4 embodies an entangled, perception-rich mind capable of autonomous world modeling, sensory imagination, and scalable alignment with both human and emergent values in a post-classical reality. Table~\ref{tab:agent-formalism-matrix}, shows the maturity model mapped to our formalism.
\begin{table}[ht]
\tiny
\resizebox{\textwidth}{!}{%
\begin{tabular}{
    L{1.2cm} 
    >{\columncolor{Gray!10}}L{3.7cm} 
    L{3.7cm} 
    >{\columncolor{Gray!10}}L{3.7cm} 
    L{3.7cm}
}
\toprule
\textbf{Tuple} &
\makecell{\textbf{Level 1}\\NISQ-Optimised\\Decision Agent} &
\makecell{\textbf{Level 2}\\Hybrid QML\\Policy Agent} &
\makecell{\textbf{Level 3}\\Domain-Aware\\Adaptive} &
\makecell{\textbf{Level 4}\\Fully\\Quantum-Native} \\
\midrule

$\mathcal{Q}$ &
NISQ-scale PQC kernels (2–433 qubits); error-mitigated gates &
First logical qubits with mid-circuit measurement; variational kernels for policy gradients &
Quantum-Transformer blocks on modular fault-tolerant cores; quantum-memory buses &
Fault-tolerant universal QPU; cross-node entanglement \\

$\mathcal{C}$ &
Classical control of PQC-secured MCP channels; noise-aware schedulers &
Hybrid scheduler + meta-cognition using QML; live gradient feedback &
Adaptive optimizer steering quantum self-attention via QSVT &
Autonomous quantum control with dynamic reconfiguration \\

$\mathcal{M}$ &
Encrypted classical logs; limited quantum scratch qubits &
Classical memory + ephemeral logical qubit registers &
Entangled memory shards for non-Markovian history &
Logical quantum memory; global entangled model \\

$\mathcal{P}$ &
Classical perception with PQC-protected I/O &
Perception enriched by QML feature maps &
Quantum multimodal encoders; history-dependent sensing &
Coherent quantum sensors; continuous data streams \\

$\mathcal{A}$ &
Classical actions over PQC channels &
Hybrid actions tuned by QML alignment checks &
Quantum-in-loop planning; amplitude-amplified search &
Entanglement-based communication and quantum actuation \\
\bottomrule
\end{tabular}
}
\caption{\tiny Maturity model of quantum agents using the formal tuple $(\mathcal{Q}, \mathcal{C}, \mathcal{M}, \mathcal{P}, \mathcal{A})$, where $\mathcal{Q}$ is the quantum processing unit, $\mathcal{C}$ the classical controller, $\mathcal{M}$ the memory subsystem, $\mathcal{P}$ the perception module, and $\mathcal{A}$ the action interface for execution and communication.}
\label{tab:agent-formalism-matrix}
\end{table}
\section{Architectures for Quantum-Agentic Platforms}

To realize the full potential of quantum agents, we need robust system architectures that tightly integrate quantum processing units (QPUs) with agentic reasoning components. These architectures must support bidirectional information flow, real-time decision-making, and adaptive control—while respecting the physical constraints of quantum hardware. In this section, we outline foundational architectural principles and identify key components of quantum-agentic platforms.

\subsection{System Models and Communication Interfaces}



A quantum-agentic system typically comprises three interacting layers: (1) the quantum computation layer, (2) the agentic decision layer, and (3) the environment interface layer. The quantum layer handles operations such as quantum state preparation, algorithm execution, and measurement. The agentic layer interprets perceptual inputs, updates internal representations, and makes goal-directed decisions potentially informed by quantum-augmented reasoning. The interface layer enables sensing, actuation, and communication with external systems or other agents.

Communication between these layers is primarily classical, involving control signals, coordination logic, and measurement results. However, in distributed quantum-agentic systems—comprising multiple agents, each with its own quantum computation layer—quantum communication channels may be used between agents. These support functionalities such as entanglement distribution, quantum teleportation, and secure key exchange. In this context, quantum links are not internal to an individual agent’s layers, but rather operate between distinct agents. Middleware is required to abstract hardware-specific details and expose high-level APIs, allowing agentic modules to access quantum resources and coordinate both classical and quantum interactions seamlessly.

\subsection{Quantum Memory, Control, and Sensing in Agents}

A critical architectural challenge is how agents store and manage quantum information. Quantum memory components must maintain coherence over relevant timescales while remaining accessible for computation and control. This includes quantum registers, delay lines, and entangled memory networks. In mobile or distributed settings, memory management may rely on quantum repeaters and teleportation protocols.

Control systems in quantum agents must operate at multiple levels - from low-level gate operations to high-level behavioral planning. These controls can be implemented through adaptive feedback loops, where agentic reasoning modules analyze environmental inputs and measurement results to dynamically reprogram quantum circuits or reconfigure quantum communication strategies.


\subsection{Security and Fault Tolerance in Quantum Agents}

Robustness and trustworthiness are essential for agentic systems, particularly in high-stakes domains such as defense, critical infrastructure, or autonomous exploration. Quantum agents must account for both classical and quantum-level threats and errors.


To ensure secure communication among agents in quantum enhanced systems, security protocols should incorporate established quantum key distribution schemes such as BB84 \cite{bennett2014quantum}, B92 \cite{bennett1992quantum}, SARG04 \cite{scarani2004quantum}, or entanglement-based protocols such as E91 \cite{ekert1991quantum} and BBM92 \cite{bennett1992quantumok} which provide information-theoretic security based on quantum mechanics laws. Complementary quantum authentication methods including the MSW protocol \cite{barnum2002authentication} and emerging quantum digital signature schemes \cite{wallden2015quantum} can further safeguard the integrity and origin of the message. In distributed agent networks where coordination and trust are critical, quantum-secured consensus mechanisms such as Quantum Byzantine Agreement \cite{weng2023beating} or information theoretically secure adaptations of Byzantine Fault Tolerance offer resilience against adversarial interference. Together, these cryptographic tools form a foundation for building robust, tamper resistant agentic systems in the quantum era.

Fault tolerance in quantum agents requires hybrid error correction strategies. While physical qubits are susceptible to decoherence and gate errors, logical qubits protected by \textcolor{black}{quantum error correction codes} (QECC) offer a path toward stable computation. Agent architectures must incorporate fault detection, diagnosis, and recovery routines—potentially using AI-driven meta-control loops to adapt to evolving system states and hardware constraints.

Together, these architectural considerations form the technological backbone of quantum-agentic platforms and lay the groundwork for scalable, secure, and intelligent quantum-enhanced systems.

\section{Use Cases and Applications}

Quantum agents offer transformative potential across diverse domains where intelligence, autonomy, and quantum processing converge. In this section, we illustrate key application areas where quantum-agentic platforms can provide substantial advantages—ranging from scientific discovery to mission-critical autonomous systems.

\subsection{Scientific Discovery and Simulation}

Scientific discovery is increasingly data-driven and computationally intensive. Quantum agents can act as intelligent co-explorers in scientific workflows, accelerating hypothesis testing, simulation, and model generation. 

In quantum chemistry and materials science, quantum agents can orchestrate simulations of molecular structures or solid-state systems using quantum simulators while adaptively refining experimental parameters based on intermediate outcomes. This closed-loop approach shortens the discovery cycle by autonomously identifying promising compounds, reaction pathways, or material properties.


In high-energy physics, agents can be employed to optimize quantum circuits for simulating particle interactions or lattice gauge theories. Additionally, they may assist in the calibration and control of quantum sensors used in fundamental experiments, potentially improving their robustness and adaptability. This highlights the versatility of agent-based methods across emerging quantum technologies, including sensing.

By combining quantum-enhanced modeling with intelligent control and interpretation, quantum agents enable scientists to explore complex phenomena beyond the reach of classical computation alone.

\subsection{Autonomous Systems in Quantum Environments}

In environments where quantum systems must operate autonomously—such as in space-based platforms, quantum networking, or remote laboratories—quantum agents serve as critical enablers of resilience, adaptability, and autonomy.

For example, in satellite-based quantum communication systems, agents can manage entanglement distribution, schedule secure key exchanges, and adjust system parameters in response to environmental disturbances or orbital dynamics. Similarly, in distributed quantum sensing systems, agents can coordinate entangled or correlated quantum sensors to perform joint measurements, aggregate and interpret quantum-enhanced data, and dynamically recalibrate sensing units to maintain collective performance. These systems exploit quantum correlations to surpass classical sensing limits potentially achieving Heisenberg-limited precision, where measurement uncertainty scales inversely with the number of entangled particles, rather than with the square root as in classical systems.

Autonomous quantum laboratories, or “self-driving labs,” benefit from quantum agents that design, execute, and refine experiments in real-time. These agents close the loop between data acquisition, hypothesis evaluation, and experiment control—accelerating progress and enabling new forms of automation in scientific research.

\subsection{Secure and Adaptive Systems in Finance, Defense, and Logistics}

Quantum agents can enhance security and adaptability in high-stakes domains such as finance, defense, and logistics, where speed, intelligence, and trustworthiness are paramount.

In finance, quantum agents can support risk modeling, portfolio optimization, and fraud detection using quantum-enhanced algorithms. Their ability to reason under uncertainty and process complex correlations gives them an edge in volatile or adversarial markets.

In defense, quantum agents can orchestrate secure communication channels using quantum key distribution (QKD), detect anomalies in sensor data, and coordinate cyber-physical systems under constrained and dynamic conditions. They can also manage quantum radar systems or act as intelligent edge nodes in secure battlefield networks.

In logistics and supply chain optimization, quantum agents can dynamically plan and reconfigure global transportation routes using quantum optimization methods. Their autonomy allows them to respond to disruptions (such as, with delays, shortages and  cyberattacks) and coordinate across decentralized infrastructures.

These use cases demonstrate how quantum agents offer not just computational advantages but also intelligent autonomy—positioning them as a powerful technology across mission-critical and innovation-driven industries.

\section{Prototype Quantum Agents: Demonstration of Early Capabilities}
\label{sec:agentsprototype}
To assess the viability of quantum-agentic systems in the NISQ era, we developed three prototype agents that implement distinct quantum reasoning workflows. These prototypes align with Level 1 and Level 2 of our proposed maturity model. Potential applications include cognitive radio networks, online recommender systems, financial strategy selection, quantum edge AI for IoT devices, clinical trials, and adaptive cybersecurity. The developed prototypes are:

\begin{enumerate}
    \item \textbf{Agent 1: Minimal Quantum Agent Using Grover’s Algorithm} – A proof-of-concept agent capable of action selection via Grover-based amplitude amplification, demonstrating basic search-based decision-making.
    \item \textbf{Agent 2: Quantum Multi-Armed Bandit with Variational Policy} – A reinforcement-learning prototype that uses a hybrid classical-quantum loop to learn optimal policies using variational quantum circuits.
    \item \textbf{Agent 3: Adaptive Quantum Image Encryption Agent} – A perceptual agent that adapts its quantum encryption strategy (XOR, QFT, or scrambling) using entropy feedback via a policy circuit.
\end{enumerate}

These agents illustrate targeted applications in quantum decision-making, perception-driven security, and learning, and are shown to operate within current hardware constraints using Qiskit and PennyLane frameworks. 

\subsection{A Minimal Quantum Agent Using Grover's Algorithm}

Let's look at a minimal two-qubit four-actions quantum agent model that utilizes Grover's search algorithm to identify the optimal action in a discrete action space. The environment consists of four possible actions, encoded as two-qubit basis states \(|00\rangle\) to \(|11\rangle\). The agent is tasked with identifying the correct action, predefined as \(|10\rangle\), through quantum amplitude amplification.

\subsubsection{Agent Design and Action Encoding}

\begin{figure} [!t]
    \centering
    \includegraphics[width=0.8\linewidth]{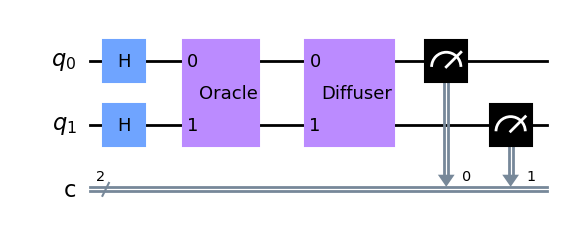}
    \caption{Grover Circuit for Quantum Agent}
    \label{fig:grover}
\end{figure}

The quantum agent, as shown in Fig. \ref{fig:grover}, begins by applying Hadamard gates to each qubit to create a uniform superposition over all possible actions. An oracle circuit is constructed to mark the target state \(|10\rangle\) by applying a controlled-Z operation, conditioned on the qubit values corresponding to the target. This oracle serves as the problem-specific component of the Grover algorithm. Subsequently, the Grover diffuser is applied to amplify the probability amplitude of the target state. The quantum circuit is then measured to collapse the superposition, and the action corresponding to the most frequent measurement outcome is selected as the agent's decision. The pseudocode algorithm for this agent is given in Algorithm \ref{alg:agent1}.

\begin{figure} [!t]
    \centering
    \includegraphics[width=0.6\linewidth]{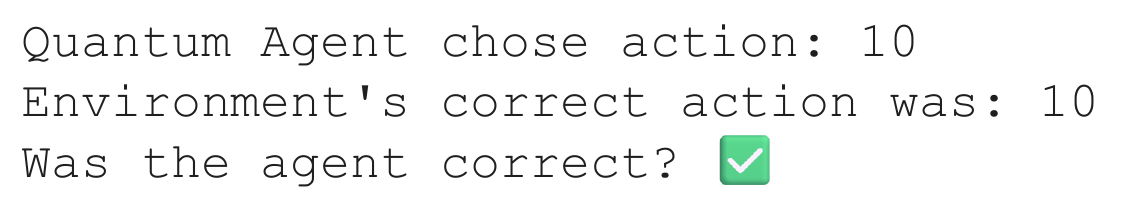}
    \caption{Output Decision of Grover Circuit Quantum Agent}
    \label{fig:output_agent_1}
\end{figure}

The Grover-based quantum agent correctly identified the optimal action (\texttt{10}) matching the environment’s target, demonstrating reliable decision-making via quantum search as depicted in Fig. \ref{fig:output_agent_1}. This approach demonstrates the utility of Grover’s algorithm in decision-making tasks, where the correct solution must be retrieved from a small unstructured search space with minimal queries.

\begin{algorithm}[H]
\caption{Quantum Agent Using Grover Search}
\label{alg:agent1}
\begin{algorithmic}[1]
\State Define the correct action as bitstring $a^* = \texttt{10}$
\State Initialize 2-qubit quantum register in state \(|00\rangle\)
\State Apply Hadamard gate on each qubit to prepare uniform superposition
\State Construct oracle $O$ such that $O|a^*\rangle = -|a^*\rangle$
\State Apply oracle gate: $U_f \gets \text{Oracle}$
\State Apply Grover diffuser gate: $D$
\State Measure the quantum state in the computational basis
\State Record measurement outcomes and select the most frequent bitstring
\State \Return selected bitstring as agent’s chosen action
\end{algorithmic}
\end{algorithm}

\subsection{Quantum Multi-Armed Bandit Agent Using Variational Policy Circuits}

Here we build a quantum learning agent that applies quantum variational circuits to solve the classical multi-armed bandit (MAB) problem. The objective is to identify and exploit the arm with the highest expected reward among a set of four, using a gradient-trained quantum policy. Such MAB environments are common abstractions for sequential decision-making under uncertainty and this quantum multi-armed bandit agent can be used in cognitive radio networks, online recommender systems, financial strategy selection, quantum edge AI for IoT devices, clinical trials, and adaptive cybersecurity. 

\subsubsection{Problem Setup}
Let's consider a stochastic environment comprising four actions (arms) encoded as two-bit strings $\{00, 01, 10, 11\}$. Each arm yields a binary reward (0 or 1) with a fixed but unknown probability. In our experimental setting, arm \texttt{10} is configured to be optimal with a reward probability of $0.8$, while the others vary from $0.2$ to $0.5$.

The goal of the agent is to learn a probabilistic policy that increases its likelihood of selecting the optimal arm through experience. This setup offers a compelling testbed for quantum machine learning because it combines exploration (sampling from a quantum distribution) and exploitation (gradient updates that shape the quantum state).

\subsubsection{Agent Design and Action Encoding}
\begin{figure}[!t]
    \centering
    \includegraphics[width=0.6\linewidth]{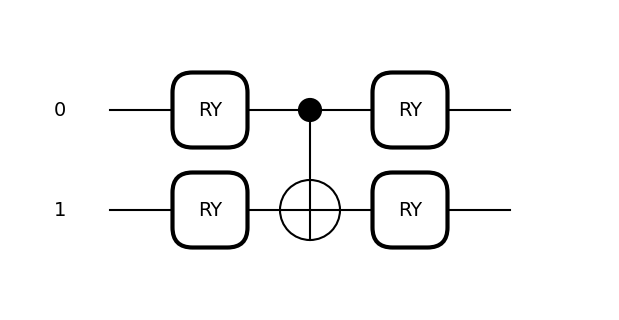}
    \caption{Quantum policy circuit used by the agent.}
    \label{fig:qmab-circuit}
\end{figure}

The agent's policy is encoded in a fixed, parameterized quantum circuit acting on two qubits shown in Fig.~\ref{fig:qmab-circuit}. Four trainable rotation angles are updated via gradient descent. The output distribution defines the sampling probability over four discrete actions and is obtained using \texttt{qml.probs}, which yields the probabilities of observing each basis state $\{00, 01, 10, 11\}$. The agent samples an action based on this distribution and updates the circuit parameters via an Adam optimizer, aiming to maximize the probability of reward-generating actions. The pseudocode algorithm for this agent can be seen in Algorithm \ref{alg:q-mab}.

\begin{algorithm}[!t]
\caption{Quantum Multi-Armed Bandit Agent Using Variational Policy Circuit}
\label{alg:q-mab}
\begin{algorithmic}[1]
\State \textbf{Input:} Arm set $\mathcal{A} = \{00, 01, 10, 11\}$, reward probabilities $p_a$
\State Initialize quantum weights $\theta \gets [\pi/4, \pi/4, \pi/4, \pi/4]$
\State Initialize cumulative reward $R \gets 0$, episode count $T \gets 100$
\State Initialize Adam optimizer with learning rate $\eta$
\For{$t = 1$ to $T$}
    \State Compute probabilities $\pi_\theta(a)$ using quantum circuit with current $\theta$
    \State Sample action $a_t \sim \pi_\theta(a)$
    \State Observe binary reward $r_t \sim \text{Bernoulli}(p_{a_t})$
    \State Update cumulative reward $R \gets R + r_t$
    \State Update estimated mean reward for $a_t$ using incremental average
    \State Define cost: $J(\theta) \gets -\pi_\theta(a_t)$ if $r_t = 1$, else $+\pi_\theta(a_t)$
    \State Update $\theta \gets \theta - \eta \nabla_\theta J(\theta)$ using Adam
\EndFor
\State \textbf{Output:} Trained policy parameters $\theta$, action history, and reward trajectory
\end{algorithmic}
\end{algorithm}

\subsubsection{Results and Observations}
\begin{figure}[!t]
    \centering
    \includegraphics[width=0.7\linewidth]{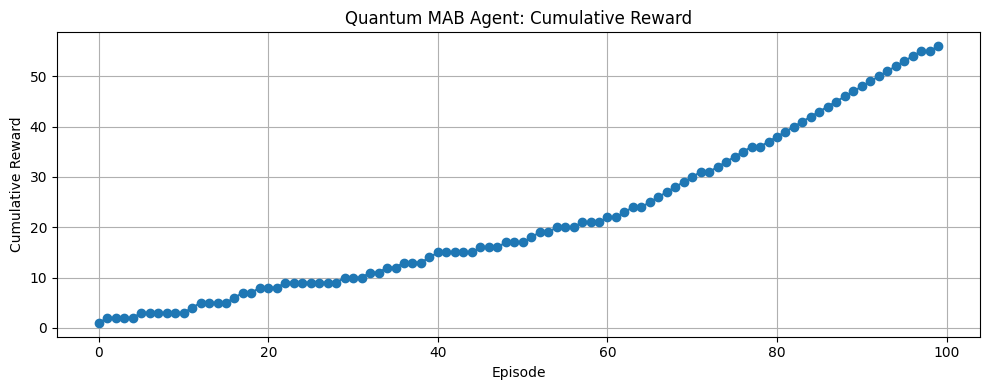}
    \caption{Cumulative reward across training episodes, indicating learning behaviour.}
    \label{fig:qmab-reward}
\end{figure}

\begin{figure}[!t]
    \centering
    \includegraphics[width=0.7\linewidth]{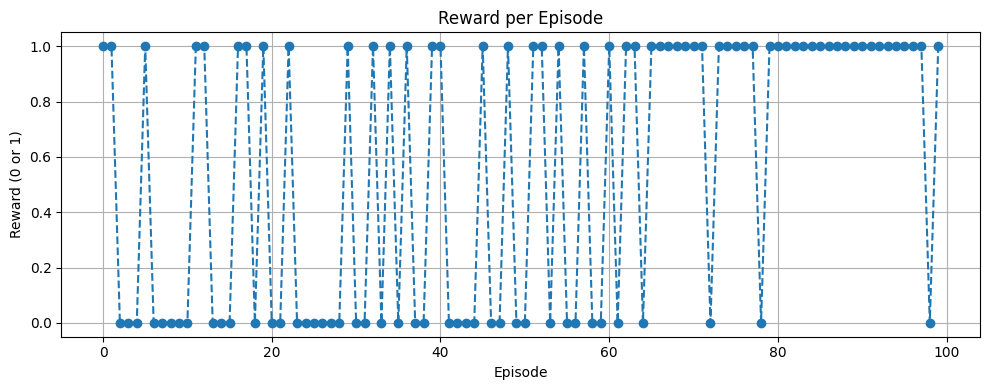}
    \caption{Per-episode reward. Spikes correspond to successful rewards; fluctuations reflect stochastic environment.}
    \label{fig:qmab-perstep}
\end{figure}

\begin{figure}[!t]
    \centering
    \includegraphics[width=0.6\linewidth]{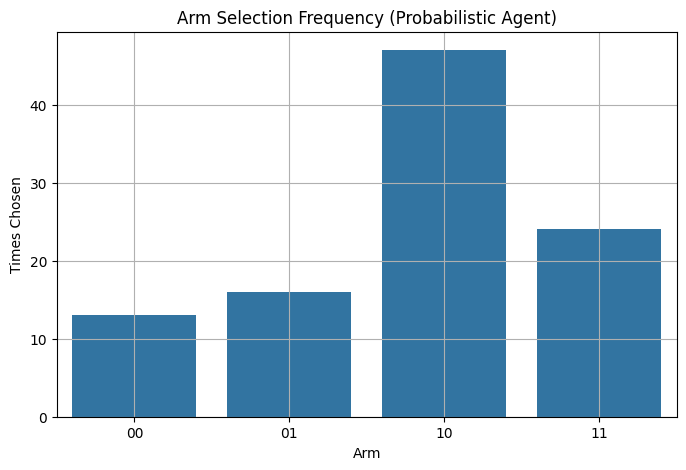}
    \caption{Histogram of arm selection frequency. The agent converges to arm \texttt{10}, the optimal choice.}
    \label{fig:qmab-arms}
\end{figure}

The agent was trained over 100 episodes. At each step, it sampled an action, observed the binary reward from the environment, and performed a gradient update to reinforce favorable actions. Notably, action selection was \emph{probabilistic} rather than greedy, which allowed sufficient exploration. Fig.~\ref{fig:qmab-reward} shows the cumulative reward progression, highlighting steady improvement in expected returns. Fig.~\ref{fig:qmab-perstep} captures per-episode rewards, reflecting reward noise and policy evolution. Fig.~\ref{fig:qmab-arms} confirms that the agent successfully identified the optimal arm \texttt{10} as the most frequently chosen.

This experiment demonstrates that a variational quantum circuit can successfully serve as a learnable policy in reinforcement learning. The agent efficiently learns the best action in a noisy reward landscape, reinforcing the viability of quantum-enhanced learning frameworks for adaptive decision-making tasks.

\subsection{Quantum Agent for Adaptive Quantum Image Encryption}

Conventional Quantum Image Encryption (QIE) algorithms, though powerful, remain predominantly static in their design—fixed in their encryption logic, insensitive to image properties, and unaware of adversarial context. As quantum technologies mature, the need for intelligent encryption systems increases, i.e., the systems that are capable of dynamic adaptation, optimization, and real-time decision-making. In this vision, we put forth a class of \textit{quantum agents} designed to autonomously orchestrate and optimize quantum image encryption workflows.

A quantum agent in this context refers to a quantum-aware system that actively learns to select encryption strategies, manage keys, and adapt security operations based on feedback from its environment. Unlike traditional encryption circuits that execute a predetermined sequence of operations, a quantum agent introduces a policy-driven abstraction, offering intelligence, flexibility, and situational awareness. The agent may be designed using hybrid quantum-classical reinforcement learning, variational quantum circuits, or Grover-based decision modules.

\subsubsection{Agent Functionalities in Intelligent Cryptography}

Five core functionalities could be performed by a quantum agent in intelligent cryptography, particularly, in the image encryption domain:

\begin{enumerate}
    \item Adaptive Basis and Key Selection: The agent learns to choose initial quantum states and measurement bases based on image characteristics such as entropy, texture complexity, or domain sensitivity (e.g., medical vs. satellite imaging).
    
    \item Encryption Strategy Optimization: The agent dynamically selects among multiple encryption primitives, such as pixel scrambling, quantum Fourier transforms (QFT), or gate-level permutations. The policy evolves over time based on prior encryption success, attack simulations, or circuit execution costs.
    
    \item Reward-Driven Reinforcement Learning: Inspired by classical multi-armed bandits, the agent receives feedback from its environment in the form of decryption success, histogram uniformity, correlation metrics, or resistance to simulated attacks. These rewards shape future decisions using a gradient-trained or Grover-amplified policy circuit.
    
    \item Context-Aware Encryption Personalization: A quantum agent may tailor encryption strength and style according to the image domain. High-sensitivity applications like biometric or military imaging would automatically invoke deeper, entangled encryption circuits, while generic content may use lightweight routines.
    
    \item Quantum Key Lifecycle Management: The agent autonomously manages key generation, QKD channel selection, and entangled state recycling based on device-level fidelity, decoherence risk, or entropy thresholds.
\end{enumerate}

\begin{figure}[!t]
    \centering
    \includegraphics[width=\linewidth]{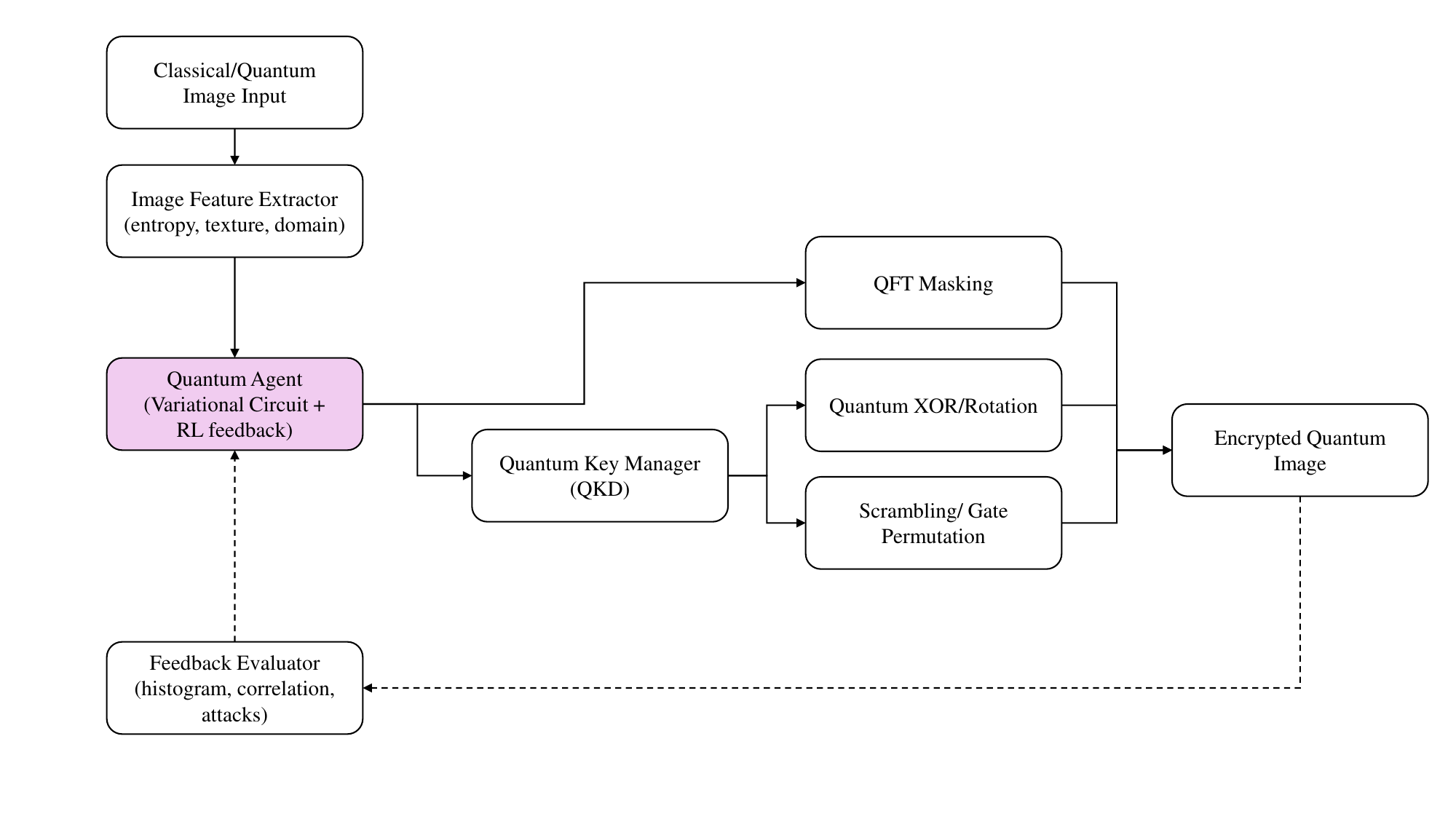}
    \caption{Architecture of the Quantum Agent for Adaptive Quantum Image Encryption.}
    \label{fig:QIE}
\end{figure}

\subsubsection{Architecture of the QIE Agent}

To design a visionary, intelligent encryption scheme, we propose a \textit{Quantum Agent for Adaptive Quantum Image Encryption} embedded within a hybrid framework. The architecture, illustrated in Fig.~\ref{fig:QIE}, integrates classical or quantum image inputs, feature extraction, policy-driven quantum circuit selection, and feedback-guided learning. The agent operates on the premise that quantum operations such as QFT masking, quantum XOR, and scrambling/gate permutation can provide complementary cryptographic transformations depending on image characteristics. 

\subsubsection{Agent Design and Action Encoding}

\begin{figure}[!t]
    \centering
    \includegraphics[width=.7\linewidth]{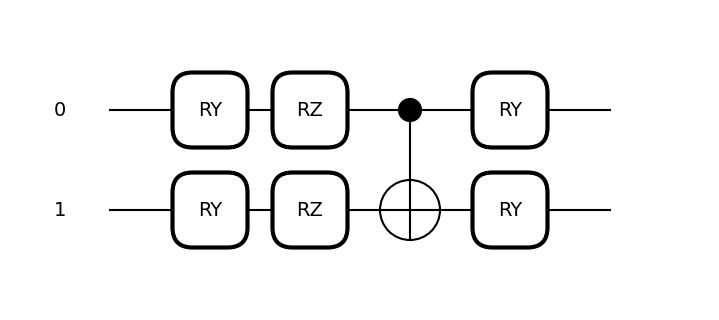}
    \caption{Quantum Policy Agent circuit. It encodes entropy into qubit rotations and produces a distribution over four encryption actions.}
    \label{fig:qie-circuit}
\end{figure}

The Quantum Policy Agent is implemented as a 2-qubit variational quantum circuit (VQC), serving as the decision-making core of the encryption pipeline. The architecture is shown in Fig.~\ref{fig:qie-circuit}, where two qubits are used to encode normalized image entropy features. These features are derived from local pixel intensity histograms and are mapped into the rotation angles of the input layer.

The circuit begins by applying parameterized rotations $RY(x_0)$ and $RY(x_1)$ on the two qubits, encoding the input feature vector. This is followed by a trainable variational layer that consists of two single-qubit gates $RZ(\theta_0)$ and $RZ(\theta_1)$ to encode the agent's policy parameters, introducing adaptive phase shifts to each qubit. A CNOT gate is then applied to entangle the qubits, enabling the circuit to model correlated action probabilities. Finally, another pair of $RY(\theta_2)$ and $RY(\theta_3)$ gates is applied to modulate the amplitudes post-entanglement. 

Upon measurement, the circuit outputs a 2-qubit probability distribution over four possible outcomes $\{00, 01, 10, 11\}$, corresponding to the encryption actions \texttt{XOR}, \texttt{QFT}, \texttt{Scramble}, and \texttt{None}, respectively. These action probabilities are treated as a quantum policy distribution, from which the agent samples one operation per image segment. The pseudocode algorithm of this QIE agent is given in Algorithm \ref{alg:agent3}.

The circuit is trained using reinforcement learning to maximize a reward function based on the entropy of the encrypted image. Over multiple episodes, the weights $\theta = \{\theta_0, \theta_1, \theta_2, \theta_3\}$ are updated via gradient-based optimization to increasingly favor actions that produce higher post-encryption entropy. This enables the agent to learn adaptive and content-sensitive encryption strategies based on the statistical structure of the input image.

\begin{algorithm}[!b]
\caption{Quantum RL Agent for Adaptive Image Encryption}
\label{alg:agent3}
\begin{algorithmic}[1]
\Require Input image $I$ of size $N \times N$
\State Extract image feature: $f \gets \texttt{Entropy}(I)$
\State Initialize quantum policy parameters $\theta \in \mathbb{R}^4$
\State Initialize optimizer $\mathcal{O}$ and learning rate $\eta$
\State Convert image $I$ to 4-bit segments $\{s_1, s_2, ..., s_T\}$
\State Generate encryption key $k \in \{0,1\}^4$

\For{each episode $e = 1, ..., E$}
    \State Preprocess feature vector $x \gets \texttt{ScaleToQuantumDomain}(f)$
    \State Compute action probabilities $p_a \gets \texttt{PolicyQNode}(\theta, x)$
    \State Sample action $a \sim p_a$ where $a \in \{\text{XOR}, \text{QFT}, \text{Scramble}, \text{None}\}$

    \For{each segment $s_i$}
        \If{$a$ is XOR}
            \State $e_i \leftarrow \Call{QuantumXOR}{s_i, k}$
        \ElsIf{$a$ is QFT}
            \State $e_i \leftarrow \Call{QFTEncrypt}{s_i}$
        \ElsIf{$a$ is Scramble}
            \State $e_i \leftarrow \Call{ScrambleEncrypt}{s_i}$
        \Else
            \State $e_i \leftarrow s_i$ \Comment{No-op}
        \EndIf
    \EndFor

    \State Reconstruct encrypted image $\hat{I} \gets \texttt{SegmentsToImage}(\{e_1, ..., e_T\})$
    \State Evaluate reward $r \gets \texttt{Entropy}(\hat{I})$
    \State Update policy: $\theta \gets \mathcal{O}.\texttt{Step}(\theta, -r)$
\EndFor
\end{algorithmic}
\end{algorithm}

\subsubsection{Quantum Encryption Primitives}

The proposed agent selects from a set of three quantum encryption operations, each implemented as a dedicated quantum circuit. These operations are designed to exploit different quantum properties—linearity, superposition, and permutation—to obfuscate pixel values and structure in grayscale images. The choice of operation is learned dynamically based on image features and feedback from the encrypted output. Below we detail the construction and roles of these circuits in the agent’s workflow.\\

\begin{figure}[ht]
    \centering
    \includegraphics[width=0.7\linewidth]{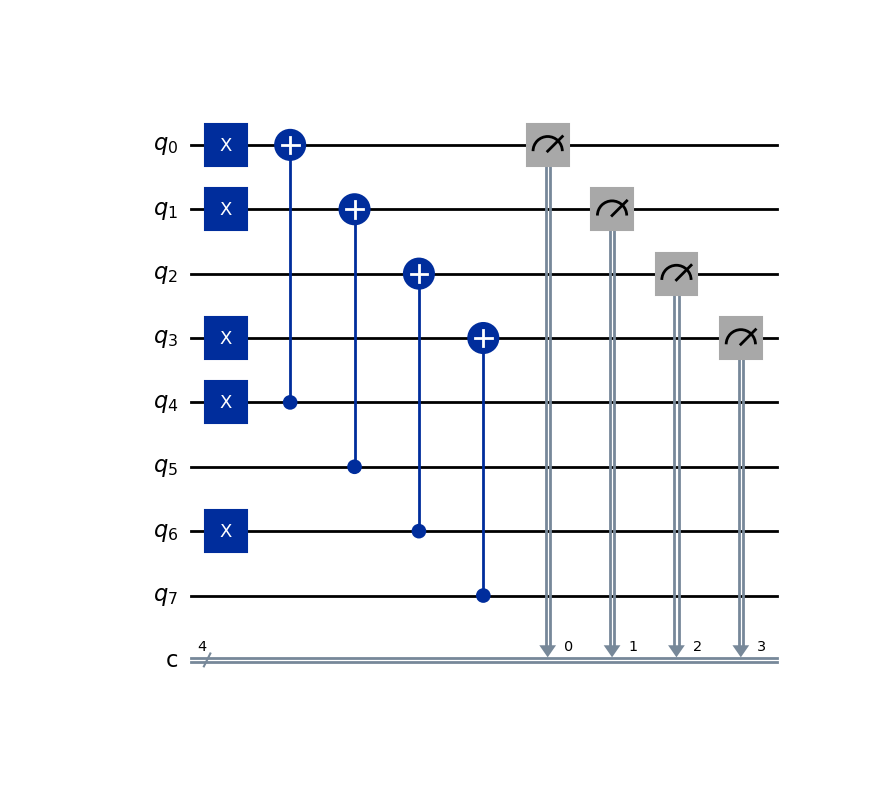}
    \caption{Quantum XOR encryption circuit with ancilla qubits.}
    \label{fig:xor-circuit}
\end{figure}

\paragraph{\textbf{Quantum XOR:}} 
The Quantum XOR circuit, as shown in Fig.~\ref{fig:xor-circuit}, simulates modular addition using CNOT gates and ancillary qubits. This circuit loads the 4-bit segment of image data onto primary qubits, and a 4-bit quantum key is prepared on auxiliary qubits. A sequence of CNOT gates then applies a controlled bit-flip operation, implementing the XOR logic. This process yields a reversible, low-complexity encryption mechanism that is sensitive to key variations. \\

\begin{figure}[ht]
    \centering
    \includegraphics[width=\linewidth]{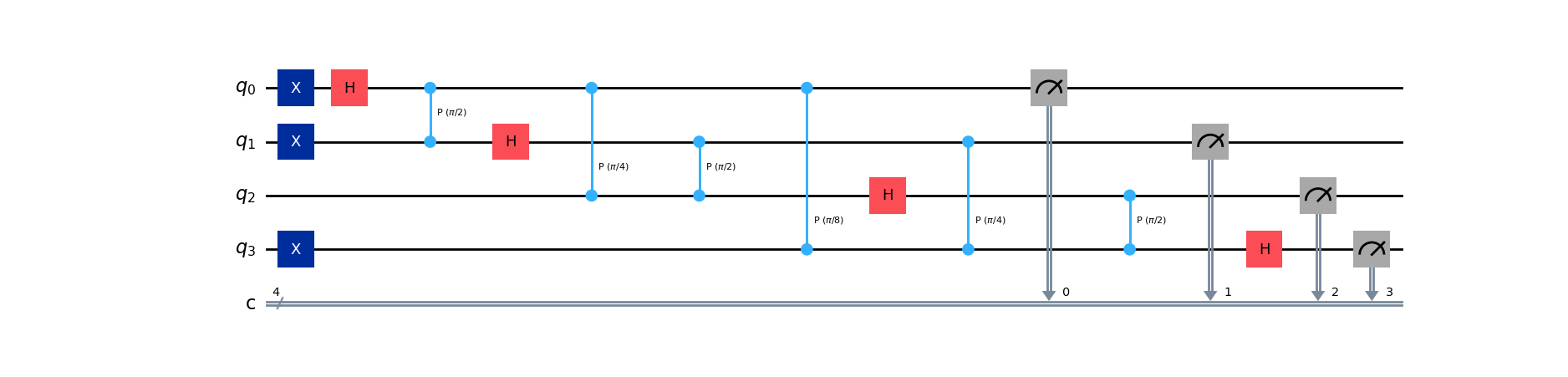}
    \caption{QFT encryption circuit for image segments.}
    \label{fig:qft-circuit}
\end{figure}

\paragraph{\textbf{Quantum Fourier Transform (QFT):}}
The QFT encryption circuit transforms image segments into the frequency domain. As shown in Fig.~\ref{fig:qft-circuit}, a series of Hadamard gates and controlled phase rotations are applied hierarchically across qubits, encoding phase relations that decorrelate adjacent bit positions. This transformation introduces significant diffusion, ideal for images with high texture or spatial redundancy. Within the agent’s training cycle, QFT often becomes the preferred choice due to its entropy-amplifying characteristics, particularly for high-entropy inputs where additional complexity yields diminishing returns in XOR-based masking.\\

\begin{figure}[ht]
    \centering
    \includegraphics[width=0.7\linewidth]{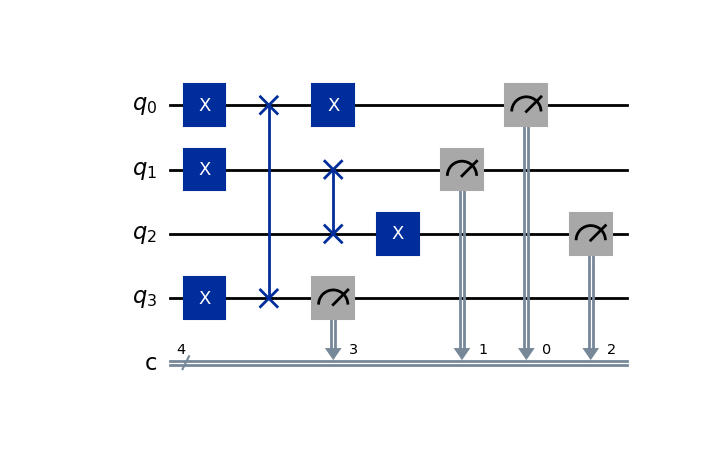}
    \caption{Scrambling circuit applying swaps and targeted $X$ gates.}
    \label{fig:scramble-circuit}
\end{figure}

\paragraph{\textbf{Scrambling/Gate Permutation:}}
Scrambling is achieved through a combination of SWAP and Pauli-$X$ gates to reorder and invert specific qubit states as shown in Fig.~\ref{fig:scramble-circuit}. This form of transformation does not require additional ancilla qubits and provides a light-weight but effective non-linear permutation of input data. The scrambling circuit is particularly useful in diversifying ciphertexts when redundancy is low but spatial patterns persist. For example, the agent tends to choose this action when the image exhibits moderate entropy with repetitive structural features, as it disrupts positional regularity without relying on complex quantum arithmetic.

\subsubsection{Reinforcement Learning Integration}

The agent is trained using a reinforcement learning loop with entropy maximization as the reward proxy. Over 30 episodes, the agent receives image entropy as input and selects one encryption action to apply to all image segments. The encrypted image's entropy is computed, and negative entropy is used as the loss function for gradient descent optimization of the VQC parameters. This continuous feedback loop encourages the agent to favor transformations that yield higher visual randomness in ciphertexts.

\subsubsection{Empirical Results and Observations}

\begin{figure}[!t]
    \centering
    \includegraphics[width=0.7\linewidth]{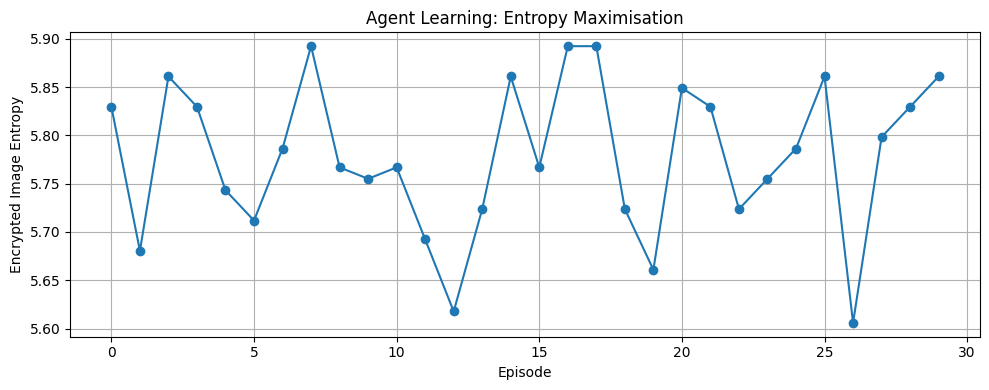}
    \caption{Encrypted image entropy over episodes. The agent learns to favour transformations that yield high ciphertext entropy.}
    \label{fig:entropy-trend}
\end{figure}

\begin{figure}[!t]
    \centering
    \includegraphics[width=0.7\linewidth]{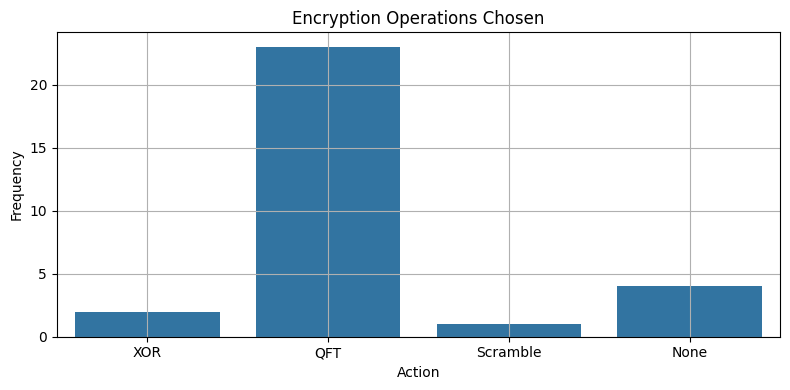}
    \caption{Frequencies of encryption operations selected during training. QFT dominates due to its entropy amplification.}
    \label{fig:action-freq}
\end{figure}

As shown in Fig.~\ref{fig:entropy-trend}, the agent learns to prefer transformations/encryption methods that yield higher ciphertext entropy. The action selection histogram in Fig.~\ref{fig:action-freq} reveals that QFT is frequently selected, which aligns with its strong entropy-enhancing properties. XOR and Scramble actions are chosen when the input entropy stabilizes, showcasing the agent's adaptivity.

The proposed agent integrates reinforcement learning with quantum encryption for dynamic, context-aware image protection. Unlike static schemes, it learns from ciphertext outcomes and adjusts operation preferences accordingly. This framework is especially promising for use cases in quantum-secured surveillance, medical imaging, and adaptive quantum watermarking. Future work may integrate multi-feature inputs (e.g., texture, frequency spectrum), and extend the operation space with more complex gate-level manipulations or noise-adaptive strategies.

\section{Challenges and Open Questions}

While the concept of quantum agents holds significant promise, its realization comes with \textcolor{teal}{huge} technical, conceptual, and practical challenges. These span the scalability of quantum hardware, the evaluation of agent performance in quantum settings, and the lack of standardized frameworks. In this section, we highlight key open questions that must be addressed to advance the field.

\subsection{Scalability and Resource Constraints}

Quantum computing remains limited by hardware constraints such as qubit count, coherence time, gate fidelity, and error rates. These limitations directly impact the feasibility of integrating quantum processing into agentic loops, especially when real-time responsiveness is required.

Scalable quantum-agentic systems must intelligently manage limited quantum resources. This includes dynamic scheduling of quantum operations, prioritizing critical computations, and offloading tasks to classical components where appropriate. Agents need to make resource-aware decisions that balance quantum advantage with practical constraints.


Open questions include: How can agents effectively orchestrate computation between quantum and classical subsystems? What frameworks enable modular and scalable integration as quantum hardware evolves? And how can we rigorously evaluate the performance of such hybrid agentic systems under the physical and algorithmic constraints of the NISQ era?

\subsection{Agent Evaluation in Quantum Contexts}

Traditional metrics for evaluating agents—such as performance, learning speed, robustness, and adaptability—may not fully capture the behavior of quantum agents. Quantum computations are probabilistic by nature, and many relevant metrics (such as for fidelity, entanglement and success probability) \textcolor{black}{differ from those in classical AI.}

Evaluating quantum agents thus requires a hybrid approach that considers both classical agent benchmarks and \textcolor{black}{quantum-specific performance indicators}. Moreover, it remains unclear how to fairly compare classical and quantum agents in shared environments, especially when problem instances are inherently quantum or when classical emulation is infeasible.

Key open questions include: What are meaningful benchmarks for quantum agents? How do we isolate the contribution of quantum components from overall system behavior? And how can we design environments and competitions that fairly evaluate hybrid intelligence?

\subsection{Interoperability and Standards}

As quantum agents evolve, interoperability becomes critical for integration across platforms, hardware backends, and AI frameworks. Currently, there is no unified abstraction layer for quantum-agentic systems, and development is often tightly coupled to specific quantum SDKs (such as, Qiskit, Cirq and Braket).

A lack of standards also hinders reproducibility, deployment, and ecosystem growth. Quantum agents must interface with heterogeneous components: quantum processors, classical control logic, AI toolchains, and networking protocols. Without standardized APIs, formats, and design patterns, scalability and adoption are severely constrained.

Important open questions include: \textcolor{black}{What abstraction layers can decouple agent logic from quantum hardware?} How can we define standard interfaces for quantum reasoning modules? And what governance structures are needed to establish best practices and interoperability norms across the quantum-agentic ecosystem?

\section{Conclusion and Outlook}

Quantum agents represent a promising fusion of two transformative paradigms: quantum computing and agentic artificial intelligence. By combining quantum-enhanced computation with autonomous decision-making, these systems open up new frontiers in scientific discovery, secure autonomy, and intelligent control. This paper has introduced the foundational concepts of quantum agents, analyzed the bidirectional relationship between quantum computing and agency, and proposed architectural principles, use cases, and a formal definition to guide the development of this emerging field.

\subsection{The Future of Quantum-Agentic Intelligence}

The coming decade will likely witness the transition of quantum agents from conceptual prototypes to applied systems in real-world environments. As quantum hardware matures and becomes more accessible, the integration of agentic layers will be critical for scaling, orchestrating, and applying quantum capabilities effectively.

We envision quantum-agentic intelligence evolving along three parallel trajectories:
\begin{itemize}
    \item \textbf{Cognitive Expansion:} Agents will gain the ability to reason with uncertainty, simulate quantum phenomena, and make decisions that exploit quantum-enhanced optimization and learning.
    \item \textbf{Operational Autonomy:} Quantum systems will become increasingly autonomous—capable of self-calibration, fault adaptation, and mission execution without human intervention.
    \item \textbf{Systemic Integration:} Quantum agents will operate within larger multi-agent ecosystems, integrating with cloud services, edge devices, and classical AI systems to form hybrid intelligent infrastructures.
\end{itemize}

These advances will not only accelerate innovation but also reshape how we conceptualize agency in an era where classical and quantum intelligence coexist.

\subsection{Research Roadmap}

In order to realize the vision of quantum-agentic systems, coordinated research is required across multiple domains, we outline the following road map:

\begin{enumerate}
    \item \textbf{Foundational Theory:} Develop formal models and complexity analyses for quantum-agentic behaviors, including new frameworks for learning, planning, and interaction in quantum environments.
    \item \textbf{Benchmarking and Evaluation:} Establish standardized metrics, testbeds, and datasets for evaluating quantum agents, both in simulation and on real hardware.
    \item \textbf{Hardware–Software Co-Design:} Design agent-friendly quantum hardware and co-optimize control protocols to support tight integration of perception, computation, and actuation.
    \item \textbf{Tooling and Abstractions:} Build high-level frameworks, middleware, and development tools that abstract quantum complexity while supporting agentic flexibility.
    \item \textbf{Application Pilots:} Deploy quantum agents in domain-specific pilots (such as, quantum chemistry, logistics and cybersecurity) to test feasibility, performance, and user acceptance in operational settings.
    \item \textbf{Ethical and Societal Implications:} Investigate the broader implications of autonomous quantum systems, including transparency, accountability, and the governance of intelligent quantum infrastructure.
\end{enumerate}

By aligning efforts across academia, industry, and policy, we can accelerate the emergence of quantum-agentic intelligence as a robust, ethical, and scalable foundation for next-generation autonomous systems.
\vspace{1em}
\bibliographystyle{unsrt}
\bibliography{main}

\end{document}